\documentclass{SCIS2025}

\usepackage{amsmath,amsfonts}
\usepackage{algorithmic}
\usepackage{graphicx}
\usepackage{textcomp}
\usepackage{xcolor}
\usepackage{textcomp}
\usepackage{array}
\usepackage{epstopdf}
\usepackage{caption}
\usepackage{multirow}
\usepackage{wrapfig}
\usepackage{setspace}
\usepackage{algorithm}
\usepackage{color, colortbl}
\usepackage{algorithmic}
\usepackage{enumitem}
\usepackage{makecell}
\usepackage{listings}
\usepackage{xcolor}
\usepackage[misc]{ifsym}
\usepackage[many]{tcolorbox}
\usepackage{fancyhdr}
\usepackage{bm}
\usepackage{booktabs}

\usepackage{url}

\usepackage{xcolor}

\newcommand{\ie}{\textit{i.e.,} }
\newcommand{\eg}{\textit{e.g.,} }

\newcommand\opt[1]{}
\newcommand\find[1]{}

\newcommand{\ls}[1]
   {\dimen0=\fontdimen6\the\font 
    \lineskip=#1\dimen0
    \advance\lineskip.5\fontdimen5\the\font
    \advance\lineskip-\dimen0
    \lineskiplimit=.9\lineskip
    \baselineskip=\lineskip
    \advance\baselineskip\dimen0
    \normallineskip\lineskip
    \normallineskiplimit\lineskiplimit
    \normalbaselineskip\baselineskip
    \ignorespaces
   }

\newenvironment{smalldescription}{
   \setlength{\topsep}{0pt}
   \setlength{\partopsep}{0pt}
   \setlength{\parskip}{0pt}
   \begin{description}
   \setlength{\leftmargin}{.2in}
   \setlength{\parsep}{0pt}
   \setlength{\parskip}{0pt}
   \setlength{\itemsep}{0pt}}{\end{description}}

\definecolor{problemblue}{RGB}{100,134,158}
\definecolor{idiomsgreen}{RGB}{0,162,0}
\definecolor{exercisebgblue}{rgb}{0,  .69,  .941}
\definecolor{deepgreen}{rgb}{0.0, 0.5, 0.0}
\definecolor{codegreen}{rgb}{0,0.6,0}
\definecolor{codegray}{rgb}{0.5,0.5,0.5}
\definecolor{codepurple}{rgb}{0.58,0,0.82}
\definecolor{backcolour}{rgb}{0.95,0.95,0.92}
\definecolor{redColor}{RGB}{255,0,0}
\definecolor{Gray}{gray}{0.1}

\definecolor{diffstart}{named}{codegreen}
\definecolor{diffincl}{named}{redColor}

\lstdefinelanguage{source}{
	language     = python,
	breaklines = true,
firstnumber=0,numberfirstline=false,columns=fullflexible,numbers=left,backgroundcolor=\color{white},
    rulecolor=\color{black}, 
    breaklines=true,sensitive=true, numbersep=5pt, xleftmargin=.015\textwidth, label=test, xrightmargin=.015\textwidth
}

\lstdefinestyle{customc}{
	belowcaptionskip=1\baselineskip,
	breaklines=false,
	frame= single,
	breaklines = true,
	xleftmargin=\parindent,
	language= Python,
	showstringspaces=false,
	basicstyle=\footnotesize\ttfamily,
	keywordstyle=\bfseries\color{green!40!black},
	commentstyle=\itshape\color{purple!40!black},
	identifierstyle=\color{blue},
	stringstyle=\color{codegreen},
	backgroundcolor=\color{gray!4}
}

\DeclareCaptionFormat{listing}{#3}
\newenvironment{findingBox}[2]{
	\begin{tcolorbox}[
colframe=black!80,
colback=gray!10,
 boxrule=.5pt,
 left=1pt,
 right = 1pt,
 top= 0pt,
 bottom=0pt,
 size=small,
 fonttitle=\bfseries,
coltitle=black,
boxrule=0.4mm,
arc=2mm
 ]{\textbf{Finding #1:} #2}
}{
	\end{tcolorbox}
}

\begin{document}
\ArticleType{RESEARCH PAPER}
\Year{2025}
\Month{January}
\Vol{68}
\No{1}
\DOI{}
\ArtNo{}
\ReceiveDate{}
\ReviseDate{}
\AcceptDate{}
\OnlineDate{}
\AuthorMark{}
\AuthorCitation{}

\title{What is Wrong with Your Code Generated by Large Language Models? An Extensive Study}{Dou S H. What is Wrong with Your Code Generated by Large Language Models? An Extensive Study}

\author[1\dag]{Shihan DOU}{}
\author[2\dag]{Haoxiang JIA}{}
\author[1]{Shenxi WU}{}
\author[1]{Huiyuan ZHENG}{}
\author[1]{\\Muling WU}{}
\author[1]{Yunbo TAO}{}
\author[1]{Ming ZHANG}{}
\author[1]{Mingxu CHAI}{}
\author[3]{Jessica FAN}{}
\author[1]{\\Zhiheng XI}{}
\author[4]{Rui ZHENG}{}
\author[5]{Yueming WU}{}
\author[5*]{Ming WEN}{mwenaa@hust.edu.cn}
\author[1*]{\\Tao GUI}{tgui@fudan.edu.cn}
\author[1]{Qi ZHANG}{}
\author[1]{Xipeng QIU}{}
\author[1*]{Xuanjing HUANG}{xjhuang@fudan.edu.cn}

\contributions{These authors contributed equally to this work.}

\address[1]{Fudan University, Shanghai 200433, China}
\address[2]{Peking University, Beijing 100871, China}
\address[3]{University of North Carolina at Chapel Hill, Chapel Hill 27599, USA}
\address[4]{Shanghai Qiji Zhifeng Co., Ltd., Shanghai 200233, China}
\address[5]{Huazhong University of Science and Technology, Wuhan 430074, China}

\abstract{The increasing development of large language models (LLMs) in code generation has drawn significant attention among researchers. To enhance LLM-based code generation ability, current efforts are predominantly directed towards collecting high-quality datasets and leveraging diverse training technologies. However, there is a notable lack of comprehensive studies examining the limitations and boundaries of existing methods. To bridge this gap, we conducted an extensive empirical study evaluating the performance of three leading closed-source LLMs and six popular open-source LLMs on three commonly used benchmarks. Our investigation, which evaluated the length, cyclomatic complexity and API number of the generated code, revealed that these LLMs face challenges in generating successful code for more complex problems, and tend to produce code that is shorter yet more complicated as compared to canonical solutions. Additionally, we developed a taxonomy of bugs for incorrect codes that includes three categories and ten sub-categories, and analyzed the root cause for common bug types. To better understand the performance of LLMs in real-world projects, we also manually created a real-world benchmark \textit{RWPB}. We analyzed bugs on \textit{RWPB} to highlight distinct differences in bug distributions between actual scenarios and existing benchmarks. Finally, we propose a novel training-free iterative method that introduces self-critique, enabling LLMs to critique and correct their generated code based on bug types and compiler feedback. Experimental results demonstrate that our approach can significantly mitigate bugs and achieve a repair success rate of 29.2\% after two iterations, indicating substantial potential for LLMs to handle more complex problems. Our comprehensive and extensive study provides insights into the current limitations of LLM-based code generation and opportunities for enhancing the accuracy and quality of the generated code.
}


\maketitle

\section{Introduction}

Code generation (or program synthesis) aims to accurately generate executable programs from problem specifications such as natural language requirement descriptions and unit test examples~\cite{le2022coderl, svyatkovskiy2020intellicode}.
Research in code generation has gained considerable attention due to its significant impact on the software industry, particularly in enhancing productivity and accessibility by accelerating program development and improving code reliability  \cite{shin2021survey, dehaerne2022code, zan2022large}.

Recently, the advancement of large language models (LLMs) has significantly propelled the field of code generation \cite{codet5, feng2020codebert, zhang2023unifying, lu2021codexglue}.
Trained on large corpora of text and code, LLMs can understand human requirements and generate the corresponding code through ``next code token prediction'' \cite{radford2019language, radford2018improving, brown2020language}.
However, the existing LLMs still struggle to generate correct code when the requirements are complex~\cite{dou2024stepcoder, roziere2023codellama, apps, starcoder}.
Despite the tremendous efforts made to enhance code generation capabilities via different means, such as using diverse pre-trained code corpora \cite{Guo2024DeepSeekCoderWT, lozhkov2024starcoder2, song2024code}, fine-tuning models on higher-quality code \cite{zheng2024opencodeinterpreter, wei2024magicoder}, or employing various natural language processing (NLP) techniques \cite{wang2022execution, yu2023mathcal, zhou2022least, wang2023two}, there remains a lack of a comprehensive understanding on the code generated by LLMs.
Thus, an extensive large-scale empirical study is demanding, which can bring the following benefits.

\textbf{First}, comprehensive multi-dimensional code analysis beyond accuracy. Accuracy (\ie passing rate of unit tests) is by far the most dominant code generation criterion, encompassing both syntactic and semantic requirements~\cite{Yuan2023NoMM, fix2009design}.
Other characteristics, such as code conciseness and complexity, are equally significant due to the importance of code maintenance~\cite{alawad2019empirical, tashtoush2023notional}.
Therefore, we expect the code generated by LLMs to be precise, concise, and low code complexity. 
\textbf{Second}, systematic bug taxonomy for LLM-generated code. Annotating the bug types in generated code and analyzing bug distributions are crucial for comprehensively understanding the limitations of LLMs in code generation. 
Specifically, previous research has primarily focused on superficial bugs in LLM-generated code, such as compilation failure and unit test failure. 
Yet, there is an absence of fine-grained categorization and investigation into the underlying causes of bugs in the generated code. 
\textbf{Third}, real-world benchmark construction and comparative analysis. There is a noticeable deficiency in benchmarks for evaluating LLMs in the real world.
The benchmarks commonly used are constructed from programming questions \cite{austin2021program, apps, chen2021evaluating} or derived from real-world code repositories and libraries \cite{zancert, li2024deveval}. 
These repositories and libraries are also often used for pre-training LLMs, which could cause data contamination and test-case leakage \cite{jain2024livecodebench, golchin2023time, weller2023according, chen2021evaluating, riddell2024quantifying, roberts2023cutoff}.
Therefore, developing a high-quality, previously untrained benchmark to evaluate LLMs under real-world conditions is crucial.
\textbf{Finally}, practical bug mitigation method with empirical validation. While identifying bugs provides valuable insights, translating these findings into actionable solutions is essential. A method that enables LLMs to systematically critique and iteratively refine their own code, guided by comprehensive bug taxonomy and compiler feedback, could significantly enhance code generation quality without requiring additional model training.

In this paper, we conduct an extensive large-scale empirical study to explore the capabilities and limitations of code generation based on LLMs. 
In particular, we thoroughly examine the correctness and complexity of the generated code by nine state-of-the-art LLMs.
We then categorize bugs, and statistically model their distribution via in-depth analysis.
To assess the effectiveness of LLMs in real-world code generation, we manually construct a high-quality benchmark, \textit{RWPB}, and analyze the bugs in code generation based on it.
Based on our findings, we propose a novel, practical method introducing self-critique to iteratively identify and fix bugs in LLM-generated code without additional training.
Our study aims to answer the following research questions:

\begin{itemize}[noitemsep, nolistsep, leftmargin=*]
    \item \textbf{RQ1: Effectiveness in Code Generation (§\ref{sec:llm-effectiveness}).} 
    What is the performance of LLMs in code generation?
    How does the complexity of tasks affect code generation?
    What are the characteristics (\eg code complexity and API number) of the generated code?
    \item \textbf{RQ2: Bugs in Code Generated by LLMs (§\ref{sec:llm-bugs}).}
    What are the different types of root causes (\ie bugs) of the LLM-generated code?
    How are these bugs distributed across widely used LLMs and popular benchmarks?
    \item \textbf{RQ3: Benchmark Construction and Evaluation (§\ref{sec:real-world-e}).}
    What is the effectiveness of LLMs in code generation on real-world projects?
    How do real-world project bugs differ from common, widely used benchmarks? 
    Do the bug distributions generated on real-world projects differ from those on existing benchmarks?
    \item \textbf{RQ4: Mitigating Bugs in Generated Code (§\ref{sec:iqgc}).}
    How to fix code bugs based on our findings without introducing additional resource consumption, such as fine-tuning LLMs.
\end{itemize}

To investigate these RQs, we collected 1,163 programming problems from three widely used benchmarks (HumanEval+~\cite{chen2021evaluating}, MBPP+~\cite{austin2021program}, and APPS+~\cite{apps}).
To perform code generation, we selected nine LLMs, including three state-of-the-art closed-source LLMs (GPT-4~\cite{gpt4tr}, GPT-3.5~\cite{gpt35}, and Claude-3 \cite{anthropic2024claude}), four well-known open-source LLMs (DeepSeek-V3~\cite{liu2024deepseek}, DeepSeek-R1~\cite{guo2025deepseek}, Llama-3-Instruct \cite{llama3} and Phi-3-Instruct \cite{abdin2024phi}), and two popular open-source code LLMs (StarCoder-2 \cite{lozhkov2024starcoder2}, DeepSeekCoder \cite{Guo2024DeepSeekCoderWT}).
We evaluated these LLMs on generated code by using unit tests. 
Experimental results show that these LLMs could achieve on average a passing rate of 48.9\%.
Specifically, DeepSeek-R1 and DeepSeek-V3 demonstrate the best performance, which is 71.2\% and 62.6\%, respectively, while Phi-3 has the lowest accuracy at 30.9\%.

Furthermore, we collected the token number of problem descriptions; the length, the cyclomatic complexity and the API number of the canonical solution. 
This investigation allows us to analyze how the length of problem descriptions and the complexity of canonical solutions affect each model's performance.
We also examined the characteristics of the generated code, focusing on factors such as code characteristics and number of comments across different LLMs and benchmarks.

Additionally, to explore the root causes of incorrect code, we conducted a comprehensive study on the bug taxonomy in such code. 
Specifically, our analysis includes two key steps: 
(1) First, we automatically created an initial taxonomy of potential bugs based on compiler feedback.
(2) Second, human experts manually examined the incorrect code and annotated the bug type.
During this phase, experts were given the flexibility to expand and refine the bug taxonomy as needed.
In particular, this process involved 22 human experts with backgrounds in software engineering, with a minimum of two experts cross-verifying each annotation.
Ultimately, we categorized the bugs into three primary and ten secondary types.
Our study reveals several interesting findings: 
\textbf{(i)} Closed-source models surpass open-source models, particularly in handling complex problems.
\textbf{(ii)} Functional bugs constitute the highest proportion, whereas syntax bugs represent the lowest. 
\textbf{(iii)} The code generated by LLMs typically has fewer lines but higher complexity and a similar API number compared to the canonical solutions.
\textbf{(iv)} Although generating more comments during code generation can enhance LLMs’ reasoning abilities, LLMs still struggle to solve more complicated problems.
\textbf{(v)} For complex problems, LLMs often fail to generate the optimal algorithms, thus leading to timeout errors.

To evaluate the efficacy of LLM-based code generation in real-world projects, we developed a real-world benchmark by manually collecting 140 programming tasks from GitHub repositories established in 2024.
Our experimental results revealed that closed-source LLMs outperform open-source LLMs in code generation quality for real-world projects, particularly excelling in reducing syntax bugs and runtime bugs. 
DeepSeek-R1 achieves the best performance with an accuracy of 77.9\%, while Phi-3 has the lowest accuracy at only 22.1\%. 
We also found that the bug distributions are inconsistent between the existing commonly used benchmarks and the real-world projects.
Our study revealed that existing LLMs are prone to introducing bugs in code generation due to their failure to capture and interpret human intent and perform the correct corresponding logic, as evidenced in both current popular benchmarks and real-world projects.
To mitigate these bugs, we propose an interactive approach involving self-critique \cite{tan2023self} without additional training costs.
Specifically, we make LLMs critique their incorrect code based on our constructed bug taxonomy and compiler feedback (\ie error messages from failed executions) to understand the problem in-depth and analyze the causes of the bug, and then re-correct the code.
This method can be performed iteratively.
Experimental results showed that our proposed method can mitigate bugs and achieve a repair success rate of 29.2\% after two iterations, which indicates substantial potential for LLMs to handle more complex tasks.
Future improvements might focus on enhancing code quality from an artificial intelligence perspective (\eg improving LLM comprehension and mitigating the misuse of function APIs by introducing explanatory comments to training code) and from a software engineering standpoint (\eg preventing syntax bugs by introducing additional syntax checkers rather than solely relying on a conventional generation paradigm in the inference time).

In summary, our paper makes the following contributions:

\begin{itemize}[noitemsep, nolistsep, leftmargin=*]
    \item \textbf{A thorough evaluation of the capabilities of LLMs in code generation.}
    We conduct extensive experiments to evaluate the performance of current widely used LLMs. 
    We also explore the effect of different types of tasks on code generation and the characteristics of generated code.
    \item \textbf{A bug taxonomy of LLM-generated code.}
    We develop a bug taxonomy based on code generated by LLMs.
    We further analyze these bugs and model their distribution to understand the root causes of inaccuracies.
    \item  \textbf{A real-world benchmark.} 
    We manually construct a real-world benchmark, \textit{RWPB}, by collecting code from the newest GitHub repositories. 
    We also analyze and compare the distribution of bugs between existing benchmarks and our real-world benchmark, highlighting the inadequacies of current models in real-world scenarios.
    \item \textbf{A novel method to mitigate code generation bugs.} 
    We propose a novel approach to mitigate bugs by introducing self-critique: allowing LLMs to find and fix bugs by iteratively critiquing their generated code.
    We also offer some suggestions and possible future solutions for improvement.
\end{itemize}

\section{Experimental Design}

\textbf{Large Language Model Selection.}
To thoroughly evaluate LLM-based code generation, we selected nine large language models of three types: open-source text LLMs, open-source code LLMs, and closed-source LLMs.
For open-source text LLMs, we selected four common models, namely DeepSeek-V3~\cite{liu2024deepseek}, DeepSeek-R1~\cite{guo2025deepseek}, Llama-3 \cite{llama3} and Phi-3 \cite{abdin2024phi}.
DeepSeek-V3 is a Mixture-of-Experts language model that achieves high performance while being significantly more cost-efficient to train and deploy.
DeepSeek-R1 is a reasoning-specialized language model that is trained through reinforcement learning and chain-of-thought training based on DeepSeek-V3.
Llama-3 is an advanced and popular LLM developed by Meta-AI, which is trained on extensive text and code corpora.
Phi-3 is another extensively used LLM, noted for its training on high-quality textbook datasets.
In the category of open-source code LLMs, our study used StarCoder-2 \cite{lozhkov2024starcoder2} and DeepSeekCoder \cite{Guo2024DeepSeekCoderWT}. 
StarCoder-2 is a widely used code model trained on 3.3 to 4.3 trillion tokens of code from The Stack v2 dataset \cite{lozhkov2024starcoder2}, encompassing over 600 programming languages.
DeepSeekCoder is another state-of-the-art open-source code LLM trained on 2 trillion code and natural language tokens.
We also selected three universally used closed-source LLMs, namely GPT-3.5 \cite{gpt35}, GPT-4 \cite{gpt4tr}, and Claude-3 \cite{anthropic2024claude}.
We set the temperature to 0.1 and use top-1 sampling to ensure consistent results. 
Additional information and details on these models are provided in Table~\ref{tab:details-llm}.

\begin{table}[t]
  \centering
  \
  \setlength\tabcolsep{6pt}
  \renewcommand{\arraystretch}{1.5}
  \caption{Overview of nine LLMs used in our study.}
\begin{tabular}{c|ccc}
\toprule
\textbf{Modality} & \textbf{Model} & \textbf{Context Window}  & \textbf{Release Date} \\
\midrule
\multirow{3}[2]{*}{Open-source/Text} & Llama-3-8B-Instruct &   8k   & Apr, 2024 \\
      & Phi-3-3.8B-Instruct & 128k    & Apr, 2024 \\
      & DeepSeek-V3 & 128k    & Dec, 2024 \\
      & DeepSeek-R1 & 128k    & Jan, 2025 \\
\midrule
\multirow{2}[2]{*}{Open-source/Code} & StarCoder-2-15B & 16k     & Feb, 2024 \\
      & DeepSeekCoder-7B &  4k     & Jan, 2024 \\
\midrule
\multirow{4}[2]{*}{Closed-source/Text}
& GPT-4-Turbo &  128k   & Dec, 2023 \\
      & GPT-3.5-Turbo-0125 &    16k   & Jan, 2024 \\
      & Claude-3 &    200k   & Mar, 2024 \\
\bottomrule
\end{tabular}
  \label{tab:details-llm}
  \vspace*{0.8em}
\end{table}

\textbf{Benchmark Selection.}
To evaluate the performance of selected LLMs and analyze bugs in code generation, we constructed a collection of assessments from well-known benchmarks.
Our selection criteria included widely recognized benchmarks to ensure relevance, the exclusion of popular libraries or projects to avoid data leakage, and diverse unit tests with high code coverage. 
The target programming language for the generated code was set to Python, as most current research focuses on improving code generation in Python \cite{Guo2024DeepSeekCoderWT, 2023opencompass, zhuo2024bigcodebench}.
Finally, we selected three widely used benchmarks, HumanEval+ \cite{chen2021evaluating}, MBPP+ \cite{austin2021program} and APPS+ \cite{apps}.
In particular, HumanEval is a popular interview Python benchmark, designed to test language comprehension, algorithmic thinking, and basic mathematics skills.
MBPP is another prevalent introductory Python benchmark created by crowd-sourced developers.
Both HumanEval+ and MBPP+ are expansions of their original versions, augmenting additional unit tests and comprising 164 and 399 programming problems, respectively \cite{liu2024your}.
APPS is a Python benchmark, including three difficulty levels (introductory, interview, and competition), collected from open-access sites, including Codewars \cite{CodeWars}, AtCoder \cite{atcoder}, Kattis \cite{Kattis}, and Codeforces \cite{codeforces}.
APPS+ is a curated version of APPS that includes manual analysis to eliminate issues such as incomplete code and syntax errors, API misuse, and missing library dependencies \cite{dou2024stepcoder}.
For APPS+, we selected 200 samples each from the categories to maintain difficulty diversity.
Our benchmark includes 164 samples from HumanEval+, 399 samples from MBPP+, and 600 samples from APPS+, totaling 1,163 samples. The average code coverage is 96.0\% as shown in Table \ref{table:maintable}.

Additionally, to evaluate the efficacy of LLM in real-world projects, we created a real-world benchmark from GitHub repositories established in 2024 (details in Section~\ref{sec:real-world-bencmark-cons}).

\textbf{Prompt Design.} 
To ensure experimental consistency across various LLMs, we used a standardized prompt format based on prior studies~\cite{liu2024your, dou2024stepcoder}. Notably, the prompt excluded library import information, requiring LLMs to autonomously identify and import necessary libraries for the APIs used. The prompt can be found in our artifact repository \cite{ar2024}.

\textbf{Computation and Labor Resources.}
To evaluate current LLMs for code generation and validate our proposed improvements, we conducted experiments on a single node with eight A100-80GB GPUs and 2TB of CPU memory, using Python 3.11. 
For code evaluation, bug analysis, and the construction of our real-world project benchmark \textit{RWPB}, we assigned 37 annotators in software engineering.
Given their countries of residence, all annotators obtained adequate compensation.

\section{LLM-Based Code Generation}

For each selected model, we generated code for programming problems from the HumanEval+ and MBPP+, and a subset of the APPS+. 
We evaluated the model's effectiveness by the passing rate of unit tests (Section~\ref{sec:llm-effectiveness}) and further analyzed the factors that influenced their performance (Section~\ref{sec:llm-factors}).

\subsection{Effectiveness of LLMs in Code Generation.}
\label{sec:llm-effectiveness}
\textbf{Performance analysis.}
Within our experimental setup, a generated code is considered correct if it passes compiler checks, runs without errors, and produces the expected outputs for predefined inputs.
The correct code must pass all unit test cases in the benchmark, as different unit test cases cover various semantics. 

\begin{table*}[t]
  \centering
  \tiny
  \setlength\tabcolsep{2pt}
  \renewcommand{\arraystretch}{2}
  \caption{
  Modalities of three commonly used benchmarks and performance of nine popular LLMs in code generation on these benchmarks. 
  Intro. and Comp. denote Introductory and Competition, respectively.
  Description denotes the token number of the description.
  LoC denotes lines of code.
  CC denotes Cyclomatic Complexity of code. 
  API denotes the number of API in the code. 
  Cov denotes branch coverage of the test cases. 
  DSC denotes DeepSeekCoder. DSV denotes DeepSeek-V3. 
  DSR denotes DeepSeek-R1.
  }
\begin{tabular}{c|l|cccccc|cccccc|ccc}
\toprule
\multicolumn{2}{c|}{\multirow{2}{*}{\textbf{Dataset}}}
& \multicolumn{1}{c}{\multirow{2}{*}{\textbf{\makecell{Description}}}}
&
& \multicolumn{1}{c}{\multirow{2}{*}{\textbf{\makecell{LoC}}}}
& \multicolumn{1}{c}{\multirow{2}{*}{\textbf{\makecell{CC}}}}
& \multicolumn{1}{c}{\multirow{2}{*}{\textbf{\makecell{API}}}}
& \multicolumn{1}{c}{\multirow{2}{*}{\textbf{\makecell{Cov}}}}
& \multicolumn{6}{|c|}{\textbf{Open-Source}}
& \multicolumn{3}{c}{\textbf{Closed-Source}}\\
\cmidrule{9-17}
\multicolumn{2}{c|}{} &       &       &       &       &       &       
& \textbf{StarCoder-2} & \textbf{DSC} & \textbf{Llama-3} & \textbf{Phi-3} & \textbf{DSV} & \textbf{DSR}
& \textbf{GPT-4} & \textbf{GPT-3.5} & \textbf{Claude-3} \\
\midrule
\multicolumn{2}{c|}{\textbf{HumanEval Plus}}
& 131.3 &  & 11.3 & 3.1 & 3.4 & 99.3\%
& 61.6\% & 58.5\% & 51.8\% & 67.1\% & 87.8\% & \textbf{93.3\%}
& 85.4\% & 64.6\% & 79.3\% \\
\midrule
\multicolumn{2}{c|}{\textbf{MBPP Plus}}
& 48.7  &  & 6.1  & 2.2 & 1.8 & 99.6\%
& 63.2\% & 66.4\% & 61.1\% & 63.2\% & 75.4\% & \textbf{81.2\%}
& 79.2\% & 73.9\% & 74.9\% \\
\midrule
\multirow{3}[6]{*}{\textbf{APPS Plus}}
& \textbf{Intro.}
& 307.7 &  & 10.3 & 3.2 & 5.6 & 94.8\%
& 42.0\% & 45.5\% & 37.5\% & 13.5\% & 57.5\% & \textbf{87.5\%}
& 72.0\% & 56.0\% & 69.5\% \\
\cmidrule{2-17}
& \textbf{Interview}
& 480.3 &  & 27.4 & 6.4 & 15.1 & 90.5\%
& 10.5\% & 20.5\% & 17.5\% & 9.0\% & 53.0\% & \textbf{66.7\%}
& 46.5\% & 25.0\% & 38.0\% \\
\cmidrule{2-17}
& \textbf{Comp.}
& 554.9 &  & 33.7 & 6.9 & 18.4 & 92.9\%
& 2.0\% & 1.0\% & 7.5\% & 1.5\% & 41.0\% & \textbf{55.0\%}
& 36.0\% & 8.5\% & 22.0\% \\
\cmidrule{2-17}
& \textbf{Average}
& 447.6 &  & 23.8 & 5.5 & 13.0 & 92.7\%
& 18.2\% & 22.3\% & 20.8\% & 8.0\% & 50.5\% & \textbf{69.8\%}
& 51.5\% & 29.8\% & 43.2\% \\
\midrule
\multicolumn{2}{c|}{\textbf{Total Average}}
& 304.6 &  & 17.8 & 4.4 & 8.9 & 96.0\%
& 35.9\% & 38.4\% & 35.1\% & 30.9\% & 62.6\% & \textbf{71.2\%}
& 63.8\% & 45.6\% & 56.7\% \\
\bottomrule
\end{tabular}
  \label{table:maintable}
  \vspace{2em}
\end{table*}

Based on Table \ref{table:maintable}, we observe a distinct performance gap across LLM architectures. DeepSeek-R1, incorporating advanced reasoning capabilities, achieves superior performance with 71.2\% overall success rate, representing a significant advancement over traditional autoregressive models. Among non-reasoning architectures, both closed-source and open-source models show varied performance, with GPT-4 (63.8\%), DeepSeek-V3 (62.6\%), and Claude-3 (56.7\%). The gap between closed-source and non-reasoning open-source models thus spans roughly 1.2\% to 32.9\%. Notably, the introduction of reasoning mechanisms in DeepSeek-R1 yields a 7.4\% - 40.3\% improvement over non-reasoning models, suggesting that architectural innovations targeting explicit reasoning capabilities, rather than mere parameter scaling, constitute a critical factor in enhancing code generation reliability.

\begin{findingBox}{1}{
The introduction of reasoning capabilities in DeepSeek-R1 yields a 7.4 - 40.3\% improvement in code generation performance over traditional models, while a persistent 1.2\% - 32.9\% performance gap exists between closed-source and open-source implementations.
\\
\textbf{Implication:} 
Software engineering research should prioritize developing reasoning-enhanced architectures and investigating the architectural factors underlying closed-source model superiority, as these represent the most promising pathways to achieving production-grade code generation rather than pursuing incremental improvements to existing approaches.
}
\end{findingBox}

\textbf{Dataset analysis.} 
The performance gap between model categories widens dramatically with increasing problem complexity. On the APPS+ Competition subset, the open-source model StarCoder-2 manages only a 2.0\% success rate, versus 36.0\% for GPT-4 and 55.0\% for DeepSeek-R1. This disparity tracks closely with rising complexity, cyclomatic complexity climbs from 3.1 (HumanEval+) to 6.9 (APPS+ Competition), and average API usage from 3.4 to 18.4, yet even the best models suffer severe performance drops. In particular, DeepSeek-R1’s reasoning enhancements yield a 19.0\% absolute improvement over GPT-4 on APPS+ Competition, underscoring that non-reasoning LLMs face fundamental limitations on complex code-generation tasks. This systematic decline in pass rates with task difficulty suggests that bugs in LLM-generated code increase predictably as problems become harder, raising critical reliability concerns for production software development.

\begin{figure}[htbp]
\centerline{\includegraphics[width=\textwidth]{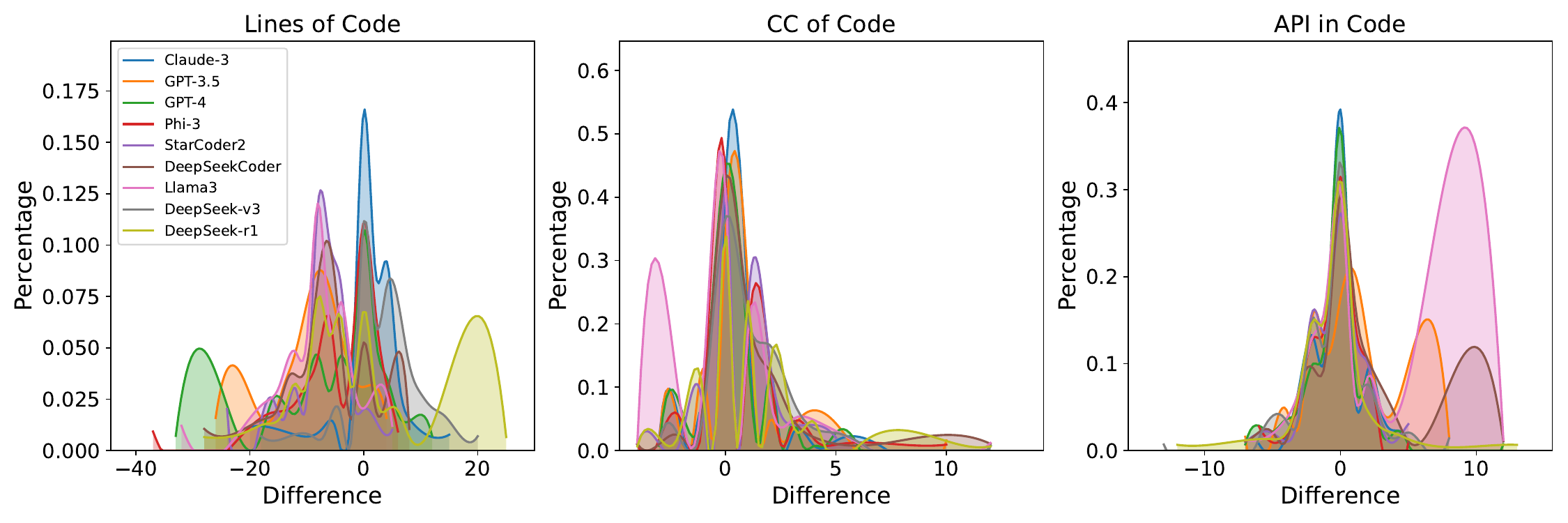}}
\caption{The differences of code characteristics between code correctly generated by models and canonical solutions on HumanEval+. CC denotes cyclomatic complexity.}
\label{fig:humanevalplus_analysis}
\vspace{0.5em}
\end{figure}

\begin{findingBox}{2}{
LLM code generation performance exhibits systematic degradation that correlates exponentially with problem complexity metrics (cyclomatic complexity, API usage, and code length), with even state-of-the-art models dropping significantly when transitioning from simple to competition-level programming tasks.
\\
\textbf{Implication:} 
Production software systems should restrict LLM-based code generation to well-bounded, low-complexity components and implement rigorous validation pipelines, as the predictable increase in bug density with complexity necessitates human oversight for any code exceeding simple algorithmic tasks.
}
\end{findingBox}

\subsection{Factors for LLMs in Code Generation}
\label{sec:llm-factors}
In this section, we analyze specific characteristics in the generated code, including the number of lines of code, cyclomatic complexity, number of APIs, and number of comment lines.

\textbf{Code size analysis.} 
We investigated the differences in code characteristics between the correct code and the canonical solution. 
Figure~\ref{fig:humanevalplus_analysis} shows the differences in lines of code, cyclomatic complexity, and API number on HumanEval+.
We can observe that the code generated by most models, including reasoning model, typically has fewer lines but slightly higher complexity, while the API number is similar to that of the canonical solutions. 
Notably, Llama3 tends to generate shorter and more API-heavy code, which is likely due to Llama3 utilizing APIs to replace parts of the complex code implementation. 
Due to the page limit, we place the experimental results for the remaining benchmarks in the open-source repository~\cite{ar2024}.

\begin{findingBox}{3}{
The code generated by the models typically has fewer lines of code but higher code complexity and a similar API number as compared to the canonical solutions.
\\
\textbf{Implication:} Optimizing code generation requires not only minimizing lines of code but also managing complexity and employing proper APIs. In practice, balancing among brevity, complexity, and API usage can lead to more maintainable and efficient solutions.
}\end{findingBox}

\begin{figure}[t]
\centerline{\includegraphics[width=1\textwidth]{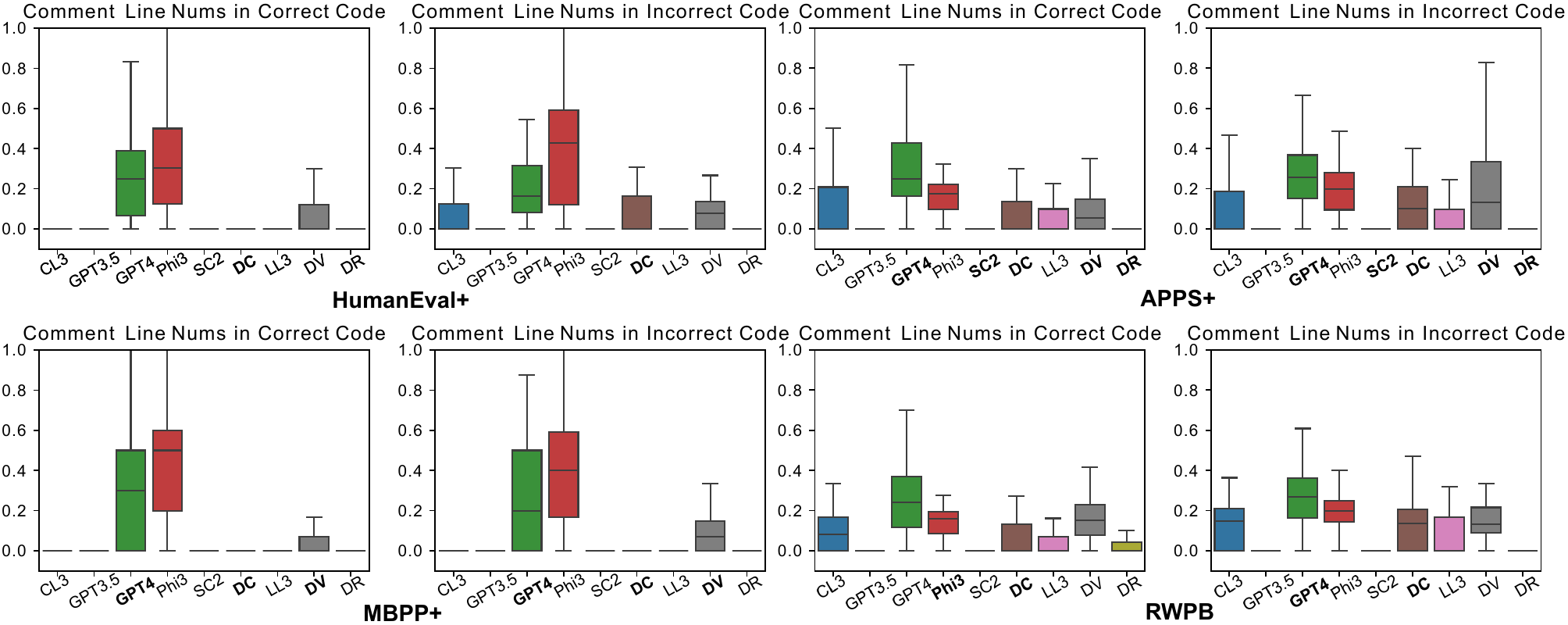}}
\caption{The difference in the comment-to-code-line ratio between the correct and incorrect code generated by LLMs. 
SC2 denotes StarCoder-2. DC denotes DeepSeekCoder. LL3 denotes Llama-3. CL3 denotes Claude-3. DSV denotes DeepSeek-V3. DSR denotes DeepSeek-R1.
The bold label indicates comments in the incorrect code are significantly (\ie $p < 0.05$) higher than in the correct code.}
\label{fig:comment}
\end{figure}

\textbf{Number of comments analysis.} Figure \ref{fig:comment} compares the number of comments in correct versus incorrect code generated by various LLMs across multiple benchmarks. Prior work has shown that introducing a chain-of-thought or “thinking” process can improve LLMs’ reasoning ability in both code reasoning and task planning tasks (e.g., \cite{imani2023mathprompter, wang2022selfconsistency, wang2022self, vacareanu2024general, wei2022chain, sharan2023llm}). Moreover, inserting comments during code generation can help models articulate their thought process and potentially enhance downstream performance \cite{song2024code, muennighoffoctopack}.

However, our results indicate a counterintuitive trend: incorrect code often contains more comments than correct code. For instance, in HumanEval+, incorrect solutions, particularly from Claude-3 and DeepSeekCoder, exhibit both a higher volume and greater variability of comments. A similar pattern emerges in APPS+, where multiple models show a higher median number of comments in incorrect code.

Two factors may contribute to this phenomenon. First, complex training examples often come with more comments, which LLMs may replicate when they are uncertain. Second, when unsure of correctness, LLMs might add explanatory comments in an attempt to clarify their logic—yet these extra comments do not necessarily yield higher accuracy. Indeed, even with more detailed commentary, the overall correctness rate remains low.

As for DeepSeek-R1, we observe a distinctly different pattern: the distribution of comments remains remarkably consistent between correct and incorrect code across all benchmarks. Unlike other models that show increased commenting in incorrect solutions, DeepSeek-R1 maintains a stable comment density regardless of solution correctness. This unique behavior can be attributed to its architectural design and training methodology. DeepSeek-R1 employs a reinforcement learning framework that explicitly separates the "thinking" process from the final code generation. During training, the model learns to internalize its reasoning chain rather than externalizing it through comments. This architectural choice fundamentally differs from traditional approaches where models are encouraged to verbalize their thought process inline with code. Consequently, DeepSeek-R1's comments serve purely as code documentation rather than as a reasoning mechanism, resulting in consistent comment patterns independent of solution quality.

In the figure, we highlight (in bold) models whose comment distributions differ significantly ($p<0.05$) between correct and incorrect code and remove outliers for clarity. These findings underscore that while comments can reflect a form of “chain of thought,” they do not inherently translate into more accurate code generation.

\begin{findingBox}{4}{
Most LLMs generate more comments in incorrect code when uncertain, but DeepSeek-R1 maintains consistent comment density regardless of correctness due to its internalized reasoning architecture.
\\
\textbf{Implication:} 
The lack of correlation between comment density and code quality in advanced models like DeepSeek-R1 indicates that future LLM architectures can achieve superior reasoning performance through internalized thought processes rather than explicit chain-of-thought verbalization, challenging current assumptions about the necessity of explanatory comments for effective code generation.}
\end{findingBox}

\section{Bugs in Code Generated by LLMs}
\label{sec:llm-bugs}
To fully understand the root causes of bugs generated by LLMs, we adopt a two-phase annotation strategy, automated annotation and manual annotation, categorizing 5,741 bugs generated from nine LLMs, identifying three primary bug types and ten secondary bug types (Section~\ref{sec:llm-taxonomy}).
We further performed a comprehensive bug analysis, including a detailed investigation into the distribution of various bugs, and discussed findings for different bug types (Section~\ref{sec:llm-analysis}).

\subsection{Taxonomy of Code Generation Bugs}
\label{sec:llm-taxonomy}

\subsubsection{Methodology} Inspired by existing work\cite{dou2024stepcoder, liu2023rltf}, we first annotated bugs in code generation into three primary bug types: \textit{Syntax Bug}, \textit{Runtime Bug}, and \textit{Functional Bug}. 
These three primary bug types, identified using the Python interpreter's output, cover all possible bug types.
Based on the primary bug types above, we adopted a two-stage process involving script-based automated annotation and manual annotation. 
The initial script-based annotation aims to efficiently annotate the primary bug types based on the results generated by the Python interpreter, enhancing the manual detailed annotation. 
The manual annotation involved experts discussing and refining the taxonomy of secondary bug types, ensuring that all bugs were correctly annotated into appropriate secondary types.

In the first stage, the script collected output from the Python interpreter during code execution and used regular expressions for initial bug annotation. 
Syntax bugs are identified by matching ``Syntax Invalid'' or ``Syntax Error'' in the error messages. 
Runtime bugs are detected by matching the ``Traceback'' field. 
Functional bugs are recognized by the ``AssertionError'' phrase, as assert statements are used to verify outputs in the experiment.
The interpreter does not provide detailed error information for ``AssertionError''. 
Therefore, the script utilized GPT-4 to generate three possible root causes along with explanations, aiding in subsequent manual analysis.

These categories align with the standard software testing hierarchy where bugs are typically classified based on when they are detected: at compile-time (syntax), runtime (execution), or through functional testing. The annotation priority ensures systematic coverage without overlap between categories.

\begin{figure*}[t]
\centerline{\includegraphics[width=0.98\textwidth]{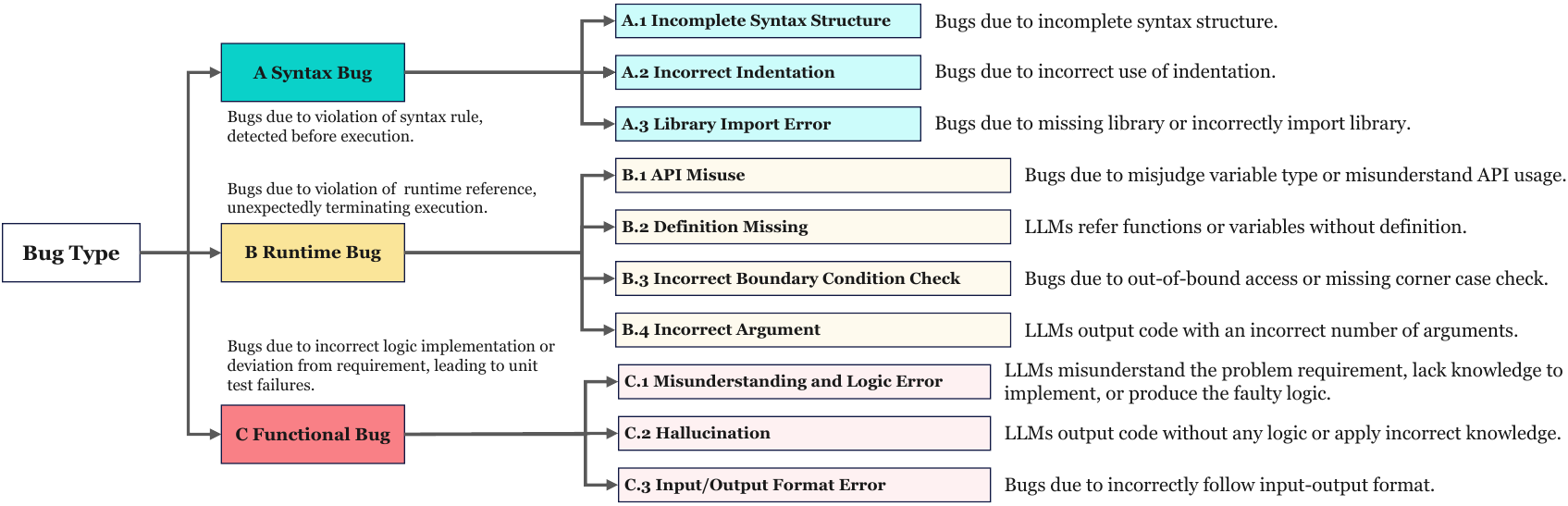}}
\caption{Taxonomy of bugs that occurred in code generated by LLMs.}
\label{fig:taxonomy}
\end{figure*}

In the second stage, 22 experts with a background in software engineering, reviewed the incorrect code annotated with primary bug types and independently developed a taxonomy for secondary bug types. 
Then, the experts combined each individual taxonomy and discussed their findings to establish a final taxonomy. 
Finally, different experts double-checked the annotation to ensure all bugs were correctly annotated.
For bugs with disagreements in annotation, all experts convened for a joint discussion until a consensus was reached. 
Additionally, experts were required to review and revise the annotation of similar bugs to ensure consistency.
Notably, because the code generation process is based on ``next code token prediction'', the generated code directly influences subsequent code generation. 
Therefore, we annotated bugs based on the root cause of the first erroneous code in the generated code.
Additionally, during the code annotation process, we found that some problems have incomplete or ambiguous functionality descriptions. 
The models attempt to infer the missing details to complete the code generation, leading to failures in unit tests. 
We refer to these issues as \textit{Ambiguous Problem Description}.
The two stages collectively took 2,400 person-hours and culminated in the classification of a taxonomy of three primary bug types and ten secondary bug types, as shown in Figure~\ref{fig:taxonomy}. 
Detailed introductions of each bug type will be discussed in the following sections.

\begin{table*}[htbp]
  \centering
  \tiny
  \setlength\tabcolsep{0.5pt}
  \renewcommand{\arraystretch}{1.3}
  \caption{Types of bugs introduced during code generation by six open-source LLMs on three widely used benchmarks (\ie HumanEval+, MBPP+, and APPS+).  
  SC2 denotes StarCoder-2. DSC denotes DeepSeekCoder. Phi3 denotes Phi-3. 
  DSV denotes DeepSeek-V3. DSR denotes DeepSeek-R1.
  All values are in \%.}
  \begin{tabular}{l|cccccc|cccccc|cccccc}
    \toprule
    \multicolumn{1}{c|}{\multirow{2}{*}{\textbf{Bug Types}}}
      & \multicolumn{6}{c|}{\textbf{HumanEval Plus}}
      & \multicolumn{6}{c|}{\textbf{MBPP Plus}}
      & \multicolumn{6}{c}{\textbf{APPS Plus}} \\
    \cmidrule{2-19}
      & \textbf{SC2} & \textbf{DSC} & \textbf{Llama3} & \textbf{Phi3} & \textbf{DSV} & \textbf{DSR}
      & \textbf{SC2} & \textbf{DSC} & \textbf{Llama3} & \textbf{Phi3} & \textbf{DSV} & \textbf{DSR}
      & \textbf{SC2} & \textbf{DSC} & \textbf{Llama3} & \textbf{Phi3} & \textbf{DSV} & \textbf{DSR} \\
    \midrule
    \rowcolor[rgb]{0.349,0.969,0.949} \textbf{A.1 Incomplete Syntax Structure}
      & 0.0 & 0.0 & 0.0 & 0.0 & 0.0 & 0.0
      & 0.3 & 0.5 & 2.4 & 1.9 & 0.0 & 0.0
      & 0.3 & 3.5 & 2.7 & 0.8 & 1.2 & 0.0 \\
    \midrule
    \rowcolor[rgb]{0.349,0.969,0.949} \textbf{A.2 Incorrect Indentation}
      & 0.0 & 0.0 & 0.0 & 0.0 & 0.0 & 0.0
      & 0.0 & 0.0 & 0.0 & 0.0 & 0.0 & 0.0
      & 0.0 & 0.2 & 0.0 & 0.0 & 0.0 & 0.0 \\
    \midrule
    \rowcolor[rgb]{0.349,0.969,0.949} \textbf{A.3 Library Import Error}
      & 3.7 & 2.4 & 10.4 & 0.0 & 0.0 & 0.0
      & 0.3 & 0.0 & 0.8 & 0.0 & 0.0 & 0.0
      & 0.0 & 0.0 & 0.5 & 0.2 & 0.0 & 0.5 \\
    \midrule
    \rowcolor[rgb]{0.039,0.820,0.788} \textbf{A Syntax Bug}
      & 3.7 & 2.4 & 10.4 & 0.0 & 0.0 & 0.0
      & 0.5 & 0.5 & 3.2 & 1.9 & 0.0 & 0.0
      & 0.3 & 3.7 & 3.2 & 1.0 & 1.2 & 0.5 \\
    \midrule
    \rowcolor[rgb]{0.988,0.945,0.792} \textbf{B.1 API Misuse}
      & 1.2 & 1.2 & 1.8 & 0.0 & 0.0 & 0.0
      & 2.9 & 2.1 & 1.3 & 2.4 & 2.7 & 1.9
      & 1.8 & 4.7 & 6.5 & 7.8 & 1.2 & 1.5 \\
    \midrule
    \rowcolor[rgb]{0.988,0.945,0.792} \textbf{B.2 Definition Missing}
      & 0.0 & 2.4 & 0.0 & 0.6 & 0.6 & 0.0
      & 0.8 & 0.3 & 0.0 & 0.0 & 1.9 & 1.3
      & 1.0 & 1.2 & 2.3 & 3.3 & 4.0 & 0.8 \\
    \midrule
    \rowcolor[rgb]{0.988,0.945,0.792} \textbf{B.3 Incorrect Boundary Condition Check}
      & 0.6 & 2.4 & 0.6 & 0.0 & 1.2 & 0.6
      & 2.4 & 1.3 & 2.1 & 1.9 & 2.1 & 2.1
      & 1.8 & 4.2 & 5.2 & 5.7 & 1.0 & 0.7 \\
    \midrule
    \rowcolor[rgb]{0.988,0.945,0.792} \textbf{B.4 Incorrect Argument}
      & 0.0 & 0.0 & 0.0 & 0.0 & 0.0 & 0.0
      & 0.0 & 0.8 & 0.0 & 0.5 & 0.0 & 0.0
      & 37.7 & 0.7 & 0.2 & 2.8 & 0.3 & 0.0 \\
    \midrule
    \rowcolor[rgb]{0.988,0.945,0.792} \textbf{B.5 Minors}
      & 3.7 & 1.8 & 1.2 & 1.2 & 0.0 & 0.0
      & 1.1 & 0.8 & 0.5 & 0.3 & 0.3 & 0.3
      & 1.0 & 1.8 & 2.7 & 1.3 & 0.0 & 1.7 \\
    \midrule
    \rowcolor[rgb]{0.980,0.898,0.600} \textbf{B Runtime Bug}
      & 5.5 & 7.9 & 3.7 & 1.8 & 1.8 & 0.6
      & 7.1 & 5.3 & 4.0 & 5.0 & 6.9 & 5.6
      & 43.3 & 12.5 & 16.8 & 21.0 & 6.5 & 4.7 \\
    \midrule
    \rowcolor[rgb]{0.992,0.827,0.835} \textbf{C.1 Misunderstanding and Logic Error}
      & 29.3 & 29.9 & 34.1 & 31.1 & 9.8 & 5.5
      & 19.0 & 18.8 & 19.8 & 20.9 & 16.1 & 12.4
      & 26.7 & 48.3 & 44.5 & 65.7 & 33.7 & 21.5 \\
    \midrule
    \rowcolor[rgb]{0.992,0.827,0.835} \textbf{C.2 Hallucination}
      & 0.0 & 1.2 & 0.0 & 0.0 & 0.0 & 0.0
      & 7.4 & 7.7 & 10.3 & 8.2 & 0.0 & 0.0
      & 5.3 & 7.5 & 10.0 & 0.5 & 0.5 & 0.0 \\
    \midrule
    \rowcolor[rgb]{0.992,0.827,0.835} \textbf{C.3 Input/Output Format Error}
      & 0.0 & 0.0 & 0.0 & 0.0 & 0.0 & 0.6
      & 2.6 & 1.3 & 1.1 & 0.5 & 0.8 & 0.0
      & 3.7 & 2.8 & 1.2 & 2.3 & 3.7 & 3.0 \\
    \midrule
    \rowcolor[rgb]{0.992,0.827,0.835} \textbf{C.4 Minors}
      & 0.0 & 0.0 & 0.0 & 0.0 & 0.6 & 0.0
      & 0.0 & 0.0 & 0.5 & 0.3 & 0.3 & 0.3
      & 1.7 & 2.0 & 2.7 & 0.7 & 1.8 & 0.5 \\
    \midrule
    \rowcolor[rgb]{0.976,0.502,0.525} \textbf{C Functional Bug}
      & 29.3 & 31.1 & 34.1 & 31.1 & 10.4 & 6.1
      & 29.1 & 27.8 & 31.7 & 29.9 & 17.2 & 12.7
      & 37.3 & 60.7 & 58.3 & 69.2 & 39.7 & 25.0 \\
    \midrule
    \rowcolor[rgb]{0.494,0.773,0.925} \textbf{D Ambiguous Problem Description}
      & 0.0 & 0.0 & 0.0 & 0.0 & 0.0 & 0.0
      & 0.0 & 0.0 & 0.0 & 0.0 & 0.0 & 0.0
      & 0.8 & 0.8 & 0.8 & 0.8 & 0.8 & 0.8 \\
    \bottomrule
  \end{tabular}
  \label{table:opensource-bugs-dv}
  \vspace{0.5em}
\end{table*}

\subsubsection{Type A: Syntax Bug.}
Syntax bugs violate the grammatical rules of the programming language being used. 
These bugs are detected by the interpreter when it tries to parse the code before execution. 
There are three secondary syntax bug types: \textit{Incomplete Syntax Structure}, \textit{Incorrect Indentation}, and \textit{Library Import Error}.

\textbf{A.1 Incomplete Syntax Structure.}
An incomplete syntax structure indicates that the generated code includes an open or partially written syntax element that has not been properly completed. 
This type of bug includes incomplete statements, unmatched parentheses, unclosed quotes, or missing colons.

\begin{lstlisting}[language=python, numbers=none, columns=fullflexible]
# unclosed Parentheses
print(" ".join(map(str, range(1, n + 1))) 
\end{lstlisting}

\textbf{A.2 Incorrect Indentation}
Python relies on indentation to define the scope of loops, conditionals, functions, and other code blocks.
Incorrect indentation violates the syntax structure, causing syntax errors or inconsistent semantics.

\textbf{A.3 Library Import Error.}
Library import allows Python programs to utilize external code, avoiding redundant development. 
Common import errors include missing import statements and incorrect import levels. 
In following example, code incorrectly imports all public functions from the \textit{heapq} library within the function body. 
However, this operation is only permitted outside the function body.
\begin{lstlisting}[language = python,numbers=none, columns=fullflexible]
def xuDRm():
    # import * only allowed at module level
    from heapq import * 
\end{lstlisting}

\subsubsection{Type B: Runtime Bug.}
\label{sec:runtime_bug}
Runtime Bugs refer to bugs that arise when the code fails to conform to runtime specifications, which are detected during execution. 
Based on the taxonomy, there are five secondary runtime bugs: \textit{API Misuse}, \textit{Definition Missing}, \textit{Incorrect Boundary Condition Check}, \textit{Incorrect Argument}, and \textit{Minors}.

\textbf{B.1 API Misuse.}
LLMs utilize APIs to enhance code execution efficiency and achieve desired functionalities. However, misinterpretation of caller attributes, incorrect API usage, or improper argument type identification can lead to API misuse, resulting in runtime errors in the generated code. The following example illustrates an attribute error arising from the misinterpretation of the variable \textit{tup}'s type.
\begin{lstlisting}[language = python,numbers=none, columns=fullflexible]
tup.sort() # 'tuple' object has no attribute sort
\end{lstlisting}

\begin{table*}[htbp]
  \centering
  \tiny
  \setlength\tabcolsep{1.5pt}
  \renewcommand{\arraystretch}{1.3}
  \caption{Types of bugs introduced during code generation by three closed-source LLMs on three widely used benchmarks (\ie HumanEval+, MBPP+, and APPS+).  
  All values are in \%.}
  \begin{tabular}{l|ccc|ccc|ccc}
    \toprule
    \multicolumn{1}{c|}{\multirow{2}{*}{\textbf{Bug Types}}}
      & \multicolumn{3}{c|}{\textbf{HumanEval Plus}}
      & \multicolumn{3}{c|}{\textbf{MBPP Plus}}
      & \multicolumn{3}{c}{\textbf{APPS Plus}} \\
    \cmidrule(lr){2-4}\cmidrule(lr){5-7}\cmidrule(lr){8-10}
      & \textbf{GPT-4} & \textbf{GPT-3.5} & \textbf{Claude-3}
      & \textbf{GPT-4} & \textbf{GPT-3.5} & \textbf{Claude-3}
      & \textbf{GPT-4} & \textbf{GPT-3.5} & \textbf{Claude-3} \\
    \midrule
    \rowcolor[rgb]{0.349,0.969,0.949} \textbf{A.1 Incomplete Syntax Structure}
      & 0.0 & 0.6 & 0.0
      & 0.0 & 0.5 & 0.0
      & 0.2 & 1.5 & 0.2 \\
    \midrule
    \rowcolor[rgb]{0.349,0.969,0.949} \textbf{A.2 Incorrect Indentation}
      & 0.0 & 0.0 & 0.0
      & 0.0 & 0.0 & 0.0
      & 0.0 & 1.2 & 0.0 \\
    \midrule
    \rowcolor[rgb]{0.349,0.969,0.949} \textbf{A.3 Library Import Error}
      & 0.6 & 2.4 & 0.0
      & 0.3 & 0.5 & 0.0
      & 1.0 & 2.0 & 2.5 \\
    \midrule
    \rowcolor[rgb]{0.039,0.820,0.788} \textbf{A Syntax Bug}
      & 0.6 & 3.0 & 0.0
      & 0.3 & 1.0 & 0.0
      & 0.2 & 4.7 & 2.7 \\
    \midrule
    \rowcolor[rgb]{0.988,0.945,0.792} \textbf{B.1 API Misuse}
      & 0.0 & 2.4 & 0.6
      & 2.4 & 1.0 & 1.8
      & 0.5 & 1.3 & 1.0 \\
    \midrule
    \rowcolor[rgb]{0.988,0.945,0.792} \textbf{B.2 Definition Missing}
      & 0.0 & 5.5 & 0.0
      & 0.0 & 0.0 & 0.0
      & 0.5 & 2.2 & 1.8 \\
    \midrule
    \rowcolor[rgb]{0.988,0.945,0.792} \textbf{B.3 Incorrect Boundary Condition Check}
      & 0.0 & 3.0 & 0.0
      & 1.5 & 0.5 & 1.5
      & 2.8 & 4.3 & 1.7 \\
    \midrule
    \rowcolor[rgb]{0.988,0.945,0.792} \textbf{B.4 Incorrect Argument}
      & 0.0 & 0.0 & 0.0
      & 1.0 & 0.0 & 0.3
      & 1.5 & 0.3 & 1.0 \\
    \midrule
    \rowcolor[rgb]{0.988,0.945,0.792} \textbf{B.5 Minors}
      & 0.6 & 0.6 & 0.6
      & 0.5 & 0.0 & 2.3
      & 1.8 & 1.7 & 2.2 \\
    \midrule
    \rowcolor[rgb]{0.980,0.898,0.600} \textbf{B Runtime Bug}
      & 0.6 & 11.6 & 1.2
      & 4.0 & 4.0 & 2.3
      & 7.2 & 9.8 & 7.7 \\
    \midrule
    \rowcolor[rgb]{0.992,0.827,0.835} \textbf{C.1 Misunderstanding and Logic Error}
      & 12.8 & 20.7 & 17.1
      & 12.0 & 15.5 & 13.8
      & 31.3 & 46.5 & 37.5 \\
    \midrule
    \rowcolor[rgb]{0.992,0.827,0.835} \textbf{C.2 Hallucination}
      & 0.6 & 0.0 & 1.2
      & 1.8 & 5.8 & 3.0
      & 7.0 & 3.0 & 4.5 \\
    \midrule
    \rowcolor[rgb]{0.992,0.827,0.835} \textbf{C.3 Input/Output Format Error}
      & 0.0 & 0.0 & 1.2
      & 2.8 & 0.8 & 2.8
      & 0.3 & 2.7 & 1.7 \\
    \midrule
    \rowcolor[rgb]{0.992,0.827,0.835} \textbf{C.4 Minors}
      & 0.0 & 0.0 & 0.0
      & 0.0 & 0.8 & 0.3
      & 0.7 & 2.7 & 2.0 \\
    \midrule
    \rowcolor[rgb]{0.976,0.502,0.525} \textbf{C Functional Bug}
      & 13.4 & 20.7 & 19.5
      & 16.5 & 22.8 & 19.8
      & 39.3 & 54.8 & 45.7 \\
    \midrule
    \rowcolor[rgb]{0.494,0.773,0.925} \textbf{D Ambiguous Problem Description}
      & 0.0 & 0.0 & 0.0
      & 0.0 & 0.0 & 0.0
      & 0.8 & 0.8 & 0.8 \\
    \bottomrule
  \end{tabular}
  \label{table:closedsource-bugs-dr}
  \vspace{2em}
\end{table*}

\textbf{B.2 Definition Missing.}
Python requires that variables and functions must be defined prior to their usage in a program. 
However, LLMs occasionally omit the definition of commonly used variables or simple functions.
As illustrated in the code below, the LLM overlooks the defining variable \textit{MOD}, which is commonly employed in algorithmic problems.
\begin{lstlisting}[language = python,numbers=none, columns=fullflexible]    
return result % MOD # variable MOD is undefined
\end{lstlisting}

\textbf{B.3 Incorrect Boundary Condition Check.}
Incorrect boundary condition check refers to the improper implementation of logic for handling the edges or limits of a range in a program. 
As shown in the code below, the program fails to verify the list length before performing a remainder operation, leading to a \textit{ZeroDivisionError} when processing an empty list.
\begin{lstlisting}[language = python,numbers=none, columns=fullflexible]    
def rotate_right(lst, n): # lst = []
    length = len(lst)
    rotation_count = n % length # ZeroDivisionError
\end{lstlisting}

\textbf{B.4 Incorrect Argument.}
LLMs occasionally disregard the specified input format in problem descriptions, resulting in mismatches with the number or type of arguments.
As shown in the code below, the problem includes two inputs: the first indicates the number of elements, and the second represents the elements to be processed. 
However, the generated code only sets one function argument to receive the inputs.
\begin{lstlisting}[language = python,numbers=none, columns=fullflexible]
-----    Input     -----
5
WEEWW
-----   Incorrect Code    -----
def min_reversal(directions): # lack an argument 
    for direction in directions:
        if direction == 'W':
\end{lstlisting}

\textbf{B.5 Minors.}
Minors in runtime bugs include \textit{Timeout Errors} and \textit{LLM-Defined Exceptions}. 
Programs exceeding this limit due to high algorithm complexity or excessive loop iterations are marked to contain a timeout error. 
LLM-defined exceptions refer to the LLM setting exceptions for conditional branches in the problem that do not have explicitly provided solutions.

\subsubsection{Type C: Functional Bug.}
Functional bug refers to the bug in the program that causes it to behave incorrectly or not as intended according to its functional requirements (\ie code runs successfully but fails to pass all unit tests). 
Based on the taxonomy, there are four secondary functional bugs: \textit{Misunderstanding and Logic Error}, \textit{Hallucination}, \textit{Input/Output Format Error}, and \textit{Minors}.

\textbf{C.1 Misunderstanding and Logic Error.}
Code generation tasks involve algorithmic problems where LLMs must extract information from natural languages and apply their knowledge to understand the requirements and establish correct logic. 
However, when faced with complex natural language descriptions, models often struggle to fully comprehend concepts, reference relationships, and conditional branches.  
As shown in the code below, LLMs incorrectly interpret integer concatenation as numerical addition. 
What is more, even if LLMs fully understand the problem description, converting this knowledge into correct logic remains challenging.
\begin{lstlisting}[language = source,numbers=none, columns=fullflexible]    
------    Problem     ------
Write a function to convert a given tuple of positive integers into a single integer.
assert tuple_to_int((1,2,3))==123
------  Incorrect Code ------
def tuple_to_int(tuple_of_ints):
    result = 0
    for digit in tuple_of_ints:
        result = result * 10 + digit
    return result
----- Canonical Solution -----
def tuple_to_int(nums):
    return int(''.join(map(str,nums)))
\end{lstlisting}

\textbf{C.2 Hallucination.}
Hallucination refers to instances where the LLM generates code that is syntactically plausible but functionally incorrect or semantically meaningless. 
As shown in the code below, the code generated by the LLM does not align with the problem requirements. 

\begin{lstlisting}[language = source,numbers=none, columns=fullflexible]    
------    Problem     ------
Write a python function to find whether a number is divisible by 11.
------  Incorrect Code ------
def is_Diff(n):
    sum = 0
    diff = 0
    while n > 0:
        digit = n % 10
        n = n // 10
        sum += digit
        diff += digit * 2 if n > 0 else digit
    return sum % 2 != 0
\end{lstlisting}

\textbf{C.3 Input/Output Format Error.}
Different from bug type B.4 (\ie incorrect argument), input/output format error refers to the incorrect order of inputs/outputs and the incorrect precision of the output data. 
As shown in the code below, the LLM incorrectly converts a float output to an integer.
\begin{lstlisting}[language = source,numbers=none, columns=fullflexible]    
------  Incorrect Code ------
def dog_age(human_years):
    ...
    return int(dog_years) # Incorrect format
----- Canonical Solution -----
def dog_age(h_age):
    ...
    return d_age
\end{lstlisting}

\textbf{C.4 Minors.}
Minors in functional bugs include \textit{Incorrect Initialization}, \textit{Sub-optimal Code}, and \textit{Infinite Loop}. 
Incorrect initialization indicates the code logic is correct, but incorrect initialization values for some variables prevent the code from passing all unit tests. 
Sub-optimal code refers to instances where LLMs generate code using sub-optimal algorithms (\eg greedy algorithms) to solve the problem, resulting in code that can only pass partial unit tests. 
Infinite loop refers to the generated code failing to meet the loop exit conditions under certain inputs, causing the code to run indefinitely.

\begin{figure*}[htpb]
\centerline{\includegraphics[width=0.96\textwidth]{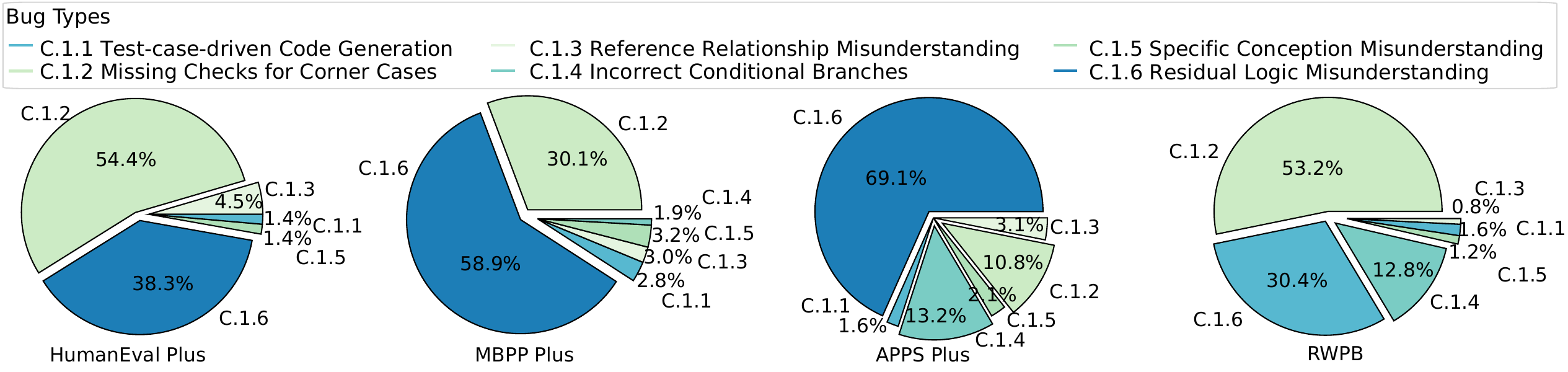}}
\caption{Distribution of misunderstanding and logic error.}
\label{fig:splitthreetype}
\end{figure*}

\subsection{Bug Analysis}
\label{sec:llm-analysis}

\textbf{Bug Distribution.} We further analyzed the distribution of primary and secondary bug types, as shown in Table~\ref{table:opensource-bugs-dv} and Table~\ref{table:closedsource-bugs-dr}. All results in the table represent the proportion of each bug type within the entire benchmark. Among the three primary bug types, syntax errors constitute the smallest proportion, whereas functional bugs are the most prevalent. This pattern holds consistently across both open-source and closed-source models, with a notable exception: DeepSeek-R1 achieves remarkably low functional bug rates (6.1\% on HumanEval+, 12.7\% on MBPP+, and 25.0\% on APPS+), significantly outperforming other models, including GPT-4 (13.4\%, 16.5\%, and 39.3\%, respectively) and Claude-3 (19.5\%, 19.8\%, and 45.7\%, respectively).
The superior performance of DeepSeek-R1 becomes more pronounced as dataset complexity increases. While most models exhibit a sharp increase in functional bugs on APPS+, with DeepSeekCoder, LLaMA-3, Phi-3, and GPT-3.5 each exceeding 50\% functional bug rates, DeepSeek-R1 maintains the lowest rate at 25\%, representing a 36\% relative improvement over GPT-4 and a 45\% improvement over Claude-3. This suggests that DeepSeek-R1 possesses enhanced semantic understanding capabilities that remain robust even when faced with complex problem descriptions.
Examining runtime bugs, DeepSeek-R1 demonstrates competitive performance with rates of 0.6\%, 5.6\%, and 4.7\% across the three benchmarks. Particularly noteworthy is its consistent avoidance of certain bug types: it produces zero hallucinations across all benchmarks, minimal incorrect arguments, and maintains low API misuse rates. This pattern indicates systematic improvements in code reliability and adherence to programming constraints.

\begin{wrapfigure}{r}{0.44\textwidth}
  \centering
  \vspace{-1.5em}
  \includegraphics[width=0.44\textwidth]{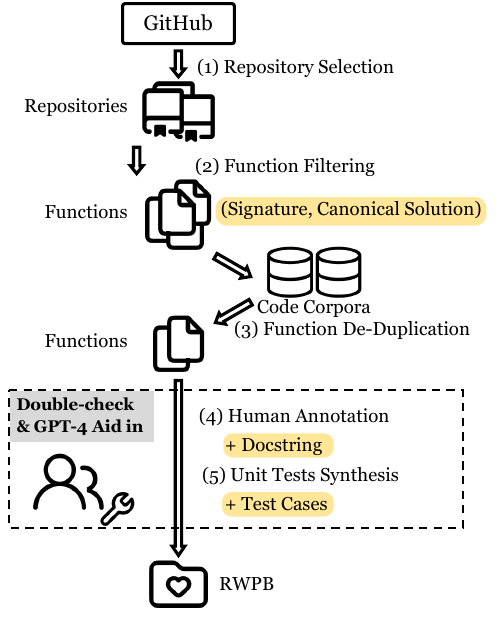} 
\caption{The process of constructing \textit{RWPB}.}
\label{fig:rwpb-process}
\end{wrapfigure}

Among all secondary bug types, misunderstanding and logic errors form the largest proportion, which is consistent across all LLMs and benchmarks. However, DeepSeek-R1 achieves the lowest rates in this category across all closed-source models: 5.5\% (HumanEval+), 12.4\% (MBPP+), and 21.5\% (APPS+), compared to GPT-4's 12.8\%, 12.0\%, and 31.3\%. This represents a 57\% reduction in logic errors on HumanEval+ and a 31\% reduction on APPS+ compared to GPT-4, suggesting that DeepSeek-R1's architecture or training methodology specifically addresses the comprehension challenges that plague other models.
We further subdivided the misunderstanding and logic error category into six labels: \textit{Test-case-driven Code Generation}, \textit{Missing Checks for Corner Cases}, \textit{Reference Relationship Misunderstanding}, \textit{Incorrect Conditional Branches}, \textit{Specific Conception Misunderstanding}, and \textit{Residual Logic Misunderstanding}, as shown in Figure \ref{fig:splitthreetype}. Test-case-driven code generation indicates LLMs generate code solely based on the test cases in the problem description, ignoring the actual functionality. Missing checks for corner cases indicate LLMs generate functionally correct code but fail to handle edge cases. Reference relationship misunderstanding indicates LLMs misinterpret reference relationships, such as numerical relationships and order of operations. Incorrect conditional branches indicate that LLMs only correctly handle some conditional branches. Specific conception misunderstanding indicates LLMs incorrectly understand specific concepts defined in the problem description. Residual logic misunderstanding refers to any remaining cases not covered by the previous labels, including instances where the understanding is correct, but the LLM fails to implement the logic correctly.

From Figure \ref{fig:splitthreetype}, missing checks for corner cases account for a large portion of errors in HumanEval+ (54.4\%) and MBPP+ (30.1\%). Meanwhile, residual logic misunderstanding is the most prevalent in APPS+, where it reaches 69.1\%, with a relatively high proportion in HumanEval+ and MBPP+. This indicates that it is more challenging for LLMs to understand problem descriptions in more complex APPS+. The exceptional performance of DeepSeek-R1, particularly its low rates of logic errors and complete absence of hallucinations, suggests that it may employ advanced reasoning mechanisms or enhanced training strategies that enable more accurate interpretation of complex problem specifications. This positions DeepSeek-R1 as a particularly promising model for code translation tasks requiring deep semantic understanding, setting a new benchmark for future LLM development in software engineering applications.

\begin{findingBox}{5}{
Functional bugs account for the highest proportion, while syntax bugs are the least common. As the complexity of the benchmark increases, the proportion of functional bugs also rises. Among secondary bug types, misunderstanding and logic errors are the most prevalent. 
\\
\textbf{Implication:} Enhancing an LLM’s ability to correctly interpret problem statements, especially corner cases and nuanced logic, represents a critical next step in code generation research. Researchers can also introduce critique techniques \cite{gero2023self} to understand or verify each code segment to improve the correctness.
}\end{findingBox}

\textbf{API Misuse.} 
API misuse accounts for the highest proportion of runtime bugs across most benchmarks and large language models (LLMs). 
There are three primary errors associated with API misuse: \textit{AttributeError} (20.9\%), \textit{TypeError} (50\%), and \textit{ValueError} (26.9\%).
An AttributeError occurs when an invalid reference is made to a missing API attribute.
A TypeError arises when an API is applied to an object of an incorrect type, while a ValueError occurs when an argument of the correct type but inappropriate value is provided. 
Thus, LLMs must accurately infer multiple factors simultaneously when invoking APIs, including the attributes of the caller, the types of API arguments, and the appropriate range of argument values.
Therefore, future work can focus on improving the LLM's inference of variable attributes, types and values to prevent API misuse.

\begin{findingBox}{6}{
API misuse represents the most common runtime bugs, necessitating accurate inference of caller attributes, argument types, and value ranges.
\\
\textbf{Implication:} Future research should enhance LLMs’ capacity to infer and validate all components involved in API calls, \ie object attributes, argument types, and permissible value ranges, to reduce runtime bugs and strengthen overall code reliability.
}\end{findingBox}

\textbf{Timeout Error.} 
A timeout error indicates that the LLM-generated code is functionally correct yet fails to complete its execution within the specified time limit, primarily due to suboptimal algorithmic choices. In complex tasks, models frequently implement nested loops or conduct deep searches without sufficient optimization or effective exit conditions, leading to excessive time complexity. The repeated use of recursive algorithms, often without appropriate bounding logic, further increases the computational cost.

\begin{findingBox}{7}{
Timeout errors are mainly caused by inefficient or unoptimized implementations, especially in complex problems where loops and recursive searches significantly increase time complexity.
\\
\textbf{Implication:} Future research should focus on LLM algorithmic optimization and complexity awareness to avoid surpassing time limits. 
}
\end{findingBox}

\section{Code Generation in Real-World Projects}
\label{sec:real-world-e}

Constructing a real-world code generation benchmark is essential for assessing LLMs' effectiveness in generating practical code. 
The distribution of bugs in real-world code may differ from that in popular benchmarks, highlighting the need for a real-world benchmark to better understand and enhance LLMs' performance. 
This section introduces the \textit{RWPB} benchmark, evaluates the performance of widely used LLMs on it, and discusses the associated challenges and limitations.

\subsection{Benchmark Construction}
\label{sec:real-world-bencmark-cons}

Our constructed benchmark \textit{RWPB} contains 140 real-world programming requirements collected from GitHub.
Each requirement includes a function signature, docstring (\ie the problem description, arguments, and returns), function body, and unit test cases.
As shown in Figure~\ref{fig:rwpb-process}, the construction process included the following five main steps:

\textbf{(1) Repository Selection.}
Currently, commonly used benchmarks are constructed from programming questions, which are disconnected from real-world program development. 
In addition, other real-world benchmarks are sourced from widely used real-world repositories and libraries.
These benchmarks suffer from data contamination and leakage due to significant overlap with the datasets used for training LLMs \cite{chen2021evaluating, riddell2024quantifying, roberts2023cutoff, golchin2023time}.
To address the shortcomings of current benchmarks, we collected Python repositories established in 2024 on GitHub.
Specifically, we focused on repositories created before April 15, 2024, and filtered out those with licenses that restricted open-source or commercial uses.
We then ranked the retained repositories in descending order of the number of stars, and ultimately obtained 600 practical repositories.

\textbf{(2) Function Filtering.}
We extracted functions from these projects and filtered out empty or initialization functions.
Following the filter approach of HumanEval \cite{chen2021evaluating}, we also removed functions that incorporated private APIs and dependencies (\ie that are not intended to be publicly accessible or used by developers), exceeded 100 lines of code, and contained many constants like strings and numbers.
In this stage, we obtained 2,307 candidate functions.

\textbf{(3) Function De-Duplication.}
Developers frequently construct their projects using existing popular tools and code repositories.
However, such code may result in data leakage since current LLMs are trained on a large fraction of GitHub and libraries.
To solve this challenge, we conduct de-duplication between our collected functions and widely used open-source code bases (\ie RedPajama Data v2, Stack v1 \cite{kocetkovstack}, Stack v2 \cite{lozhkov2024starcoder2}, RefinedWeb \cite{penedo2023refinedweb}, and the Pile \cite{gao2020pile}) by utilizing the MinHash algorithm \cite{broder1997resemblance}.
This method significantly reduces the potential for data leakage in our benchmark.
Finally, we retained 316 candidate functions.

\textbf{(4) Human Annotation.}
After eliminating potentially leaked functions, we annotated the docstring for these functions.
We engaged nine experts to manually annotate each function's docstring, including functionality descriptions, arguments, and returns.
The annotated docstring follows three criteria: naturalness, accuracy, and clarity.
Specifically, firstly, the docstring should follow the human writing style, making it naturally readable from a developer’s perspective. 
Secondly, the description should accurately describe the function's functionality.
The arguments and return values also should be accurate, such as the type and name should be identical to the code.
Finally, the docstring should be clear, precise, and concise, avoiding any redundant and irrelevant information.
Each docstring involved a dual annotation process and three annotators: one expert manually wrote a docstring, and the others provided a meticulous double-check.
If the two experts responsible for the double-check provided differing results, the drafted docstring would be manually refined until a consensus was reached.
Experts also eliminated functions that exhibited a conglomeration of functionalities and were difficult to summarize exactly and thoroughly. 
Finally, this stage retained 140 functions, each containing an accurate and comprehensive docstring.

\textbf{(5) Unit Tests Synthesis.}
We engaged six experts to write unit test cases for each function.
The generation of test cases is based on the following key criteria:
Each function contains at least three test cases; these test cases are ensured to cover the entire code of the function body; and they comprise regular and boundary cases.
In addition, similar to Stage Four, we implemented a double-check synthesis process. 
For each function, one expert was responsible for writing test cases.
Subsequently, two experts review all samples to ensure the coverage of the unit tests is 100\%.

Finally, we selectively retained 140 functions to construct our real-world benchmark \textit{RWPB}.
Each function contains a function signature, an accurate and comprehensive docstring, a canonical solution, and an average of 4.9 test cases.

\begin{table}[t]
  \centering
  \footnotesize
  \setlength\tabcolsep{5pt}
  \renewcommand{\arraystretch}{0.8}
  \caption{Types of bugs introduced during code generation by nine popular LLMs on our constructed real-world benchmark \textit{RWPB}. DSV denotes DeepSeek-V3. DSR denotes DeepSeek-R1. All values are in \%.}
  \label{table:rwpb}
  \vspace{2mm}
\begin{tabular}{l|cccccc|ccc}
\toprule
\multicolumn{1}{c|}{\multirow{2}{*}{\textbf{Bug Types}}}
  & \multicolumn{6}{c|}{\textbf{Open-Source}}
  & \multicolumn{3}{c}{\textbf{Closed-Source}} \\
\cmidrule{2-10}
  & \textbf{StarCoder2} & \textbf{DeepSeekCoder} & \textbf{Llama3}
  & \textbf{Phi3} & \textbf{DSV} & \textbf{DSR}
  & \textbf{GPT4}  & \textbf{GPT3.5} & \textbf{Claude3} \\
\midrule
\rowcolor[rgb]{0.188,0.878,0.459} \textbf{PASS} & 31.4 & 32.9 & 22.9 & 22.1 & 75.7 & 77.9 & 44.3 & 34.3 & 45.7 \\
\midrule
\rowcolor[rgb]{0.349,0.969,0.949} \textbf{A.1}  & 0.0  & 0.0  & 0.7  & 0.0  & 0.0  & 0.0  & 0.7  & 0.7  & 0.0  \\
\midrule
\rowcolor[rgb]{0.349,0.969,0.949} \textbf{A.2}  & 2.1  & 0.0  & 0.7  & 2.9  & 0.0  & 0.0  & 2.1  & 2.1  & 0.0  \\
\midrule
\rowcolor[rgb]{0.349,0.969,0.949} \textbf{A.3}  & 4.3  & 2.9  & 2.9  & 2.1  & 1.4  & 2.1  & 2.9  & 4.3  & 1.4  \\
\midrule
\rowcolor[rgb]{0.039,0.820,0.788}   \textbf{A}    & 6.4  & 2.9  & 4.3  & 5.0  & 1.4  & 2.1  & 5.7  & 7.1  & 1.4  \\
\midrule
\rowcolor[rgb]{0.988,0.945,0.792} \textbf{B.1}  & 7.9  & 7.1  & 10.7 & 10.0 & 7.9  & 3.6  & 2.9  & 5.0  & 2.9  \\
\midrule
\rowcolor[rgb]{0.988,0.945,0.792} \textbf{B.2}  & 1.4  & 0.7  & 0.7  & 5.7  & 0.0  & 0.0  & 1.4  & 2.1  & 7.9  \\
\midrule
\rowcolor[rgb]{0.988,0.945,0.792} \textbf{B.3}  & 5.0  & 4.3  & 3.6  & 2.1  & 0.7  & 0.7  & 1.4  & 3.6  & 2.9  \\
\midrule
\rowcolor[rgb]{0.988,0.945,0.792} \textbf{B.4}  & 12.9 & 19.3 & 19.3 & 16.4 & 0.0  & 1.4  & 7.1  & 12.1 & 7.1  \\
\midrule
\rowcolor[rgb]{0.988,0.945,0.792} \textbf{B.5}  & 0.0  & 1.4  & 3.6  & 2.9  & 0.0  & 0.0  & 0.0  & 1.4  & 1.4  \\
\midrule
\rowcolor[rgb]{0.980,0.898,0.600}   \textbf{B}    & 27.1 & 32.9 & 37.9 & 37.1 & 8.6  & 5.7  & 12.9 & 24.3 & 22.1 \\
\midrule
\rowcolor[rgb]{0.992,0.827,0.835} \textbf{C.1}  & 27.1 & 21.4 & 24.3 & 28.6 & 12.9 & 11.4 & 26.4 & 25.7 & 25.0 \\
\midrule
\rowcolor[rgb]{0.992,0.827,0.835} \textbf{C.2}  & 6.4  & 5.7  & 7.1  & 4.3  & 0.0  & 0.0  & 5.0  & 6.4  & 0.0  \\
\midrule
\rowcolor[rgb]{0.992,0.827,0.835} \textbf{C.3}  & 1.4  & 2.1  & 0.7  & 0.7  & 1.4  & 2.1  & 2.1  & 1.4  & 3.6  \\
\midrule
\rowcolor[rgb]{0.992,0.827,0.835} \textbf{C.4}  & 0.0  & 2.1  & 2.9  & 2.1  & 0.0  & 0.7  & 3.6  & 0.7  & 2.1  \\
\midrule
\rowcolor[rgb]{0.976,0.502,0.525}   \textbf{C}    & 35.0 & 31.4 & 35.0 & 35.7 & 14.3 & 14.3 & 37.1 & 34.3 & 30.7 \\
\bottomrule
\end{tabular}
\vspace{0.5em}
\end{table}

\subsection{Effectiveness in \textit{RWPB} and Bugs Analysis}

\textbf{Effectiveness.}
As shown in Table~\ref{table:rwpb}, the evaluated LLMs demonstrate significant performance variation. DeepSeek-R1 and DeepSeek-V3 achieve the highest passing rates at 77.9\% and 75.7\%, respectively, substantially outperforming all other models. Among the remaining closed-source LLMs, Claude-3 reaches 45.7\% and GPT-4 achieves 44.3\%, while GPT-3.5 lags behind at 34.3\%. Open-source models show notably lower performance, with StarCoder-2 and DeepSeekCoder achieving passing rates of 31.4\% and 32.9\%, respectively, while Llama-3 and Phi-3 trail at 22.9\% and 22.1\%.

A key observation emerges from analyzing the distribution of bug types across models. DeepSeek-R1 and DeepSeek-V3 excel particularly in avoiding runtime bugs, with occurrence rates of only 5.7\% and 8.6\% respectively, significantly lower than the 12.9\%-37.9\% range observed in other models. This superior performance suggests these models have developed a more robust understanding of execution semantics. Notably, both DeepSeek models demonstrate exceptional capability in handling Type B.4 bugs, which typically involve complex control flow issues. When excluding the high-performing DeepSeek models, a clear pattern emerges: traditional open-source models exhibit higher rates of runtime bugs (27.1\%-37.9\%) compared to their closed-source counterparts (12.9\%-24.3\%), indicating that closed-source LLMs generally possess better runtime semantic understanding.

In contrast, the performance gap narrows considerably for functional bugs. Most models cluster within a similar range of 30.7\% to 37.1\%, though DeepSeek-V3 and DeepSeek-R1 maintain their advantage with only 14.3\% occurrence. Further investigation into the overlap of functional bugs across LLMs revealed limited commonality, suggesting that different models possess distinct areas of expertise in real-world scenarios. This variability likely stems from differences in training data sources, model architectures, training methodologies, and hyperparameter configurations, which collectively shape each model's unique understanding and capabilities.

\begin{findingBox}{8}{
Closed-source LLMs outperform open-source LLMs in generated code quality for real-world projects, particularly excelling in reducing syntax and runtime bugs. 
\\
\textbf{Implication:} Researchers should incorporate a diverse range of real projects into the training set to generate the generalization of LLMs on real-world benchmarks.
The substantial performance gains observed in reasoning-enhanced models suggest that future research should explore hybrid training approaches that combine supervised learning on large-scale code corpora with reinforcement learning from compiler feedback, test execution results, and static analysis tools.
}\end{findingBox}

\textbf{Bug Analysis.}
We analyzed the bug distribution in the real-world project benchmark, \textit{RWPB}, and compared the distribution with HumanEval+, MBPP+, and APPS+.
Similar to the existing benchmarks, the proportion of syntax bugs in \textit{RWPB} is the lowest, indicating that LLMs can effectively avoid syntax bugs in real-world scenarios as well. 
Furthermore, Figure \ref{fig:comment} demonstrates that the number of comments between correct and incorrect code in \textit{RWPB} exhibits the same trend as the existing benchmarks: incorrect code contains more comments than correct code. 
What is more, the distribution of misunderstanding and logic errors is similar. 
Figure \ref{fig:splitthreetype} shows that missing checks for corner cases are the most prevalent sub-categories (53.2\%), followed by the residual logic misunderstanding (30.4\%).
However, unlike existing benchmarks, the proportion of runtime bugs is higher for all LLMs, particularly for incorrect arguments and incorrect boundary condition checks.

\begin{findingBox}{9}{
For \textit{RWPB}, LLMs align with existing benchmarks in most parts. 
However, the proportion of runtime bugs, especially incorrect arguments and incorrect boundary condition checks, is significantly higher compared to existing benchmarks.
\\
\textbf{Implication}: Encouraging the model to perform self-critique, such as checking the input format aligned with the requirement, can improve its performance.
}\end{findingBox}

\section{Mitigating Bugs in Generated Code}
\label{sec:iqgc}

In the preceding section, we analyzed and categorized the types of bugs in the LLM-generated code, yielding several interesting findings. 
Building on these insights, we propose a straightforward and effective method that enables LLMs to self-critique their generated code and iteratively fix bugs.

\subsection{Method}

Recent studies \cite{dong2022survey, wies2023learnability} have demonstrated that providing LLMs with more detailed information about specific tasks can enhance their performance on those tasks, such as in math reasoning \cite{ahn2024large}. 
Additionally, LLMs can leverage the world knowledge acquired during the pre-training phase to critique text, aiding humans in identifying errors in natural language.
For instance, Saunders et al. \cite{saunders2022self} fine-tuned LLMs to generate natural language critiques to help humans find flaws in a topic-based summarization task.
Wang et al. \cite{tan2023self} also introduced critique data in the fine-tuning phase to reduce the harmful responses generated by LLMs.
Inspired by these findings, we introduce a self-critique method to enable LLMs to analyze and critique their own generated code, to identify hidden bugs, and subsequently correct them.

\begin{figure}[t]
\centerline{\includegraphics[width=0.6\textwidth]{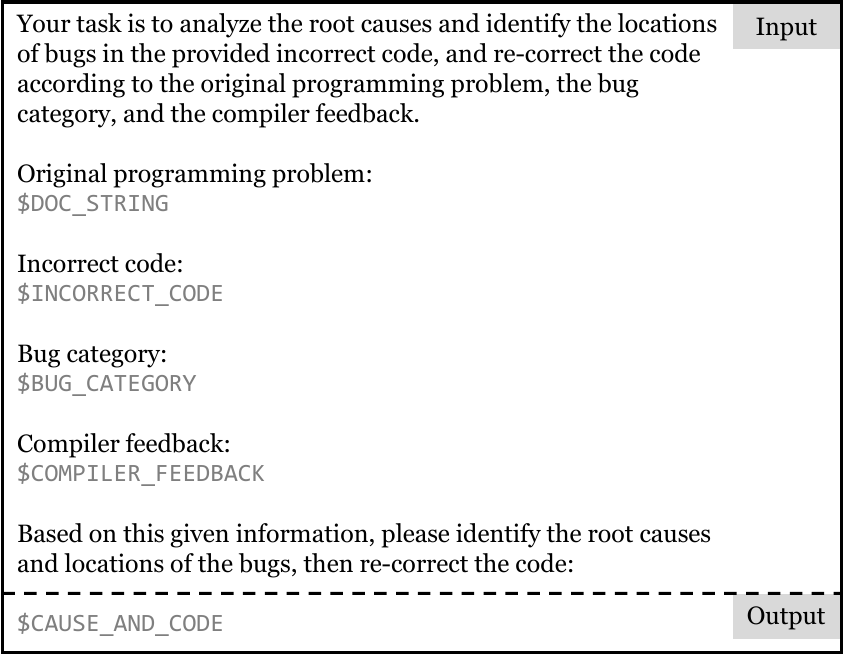}}
\caption{The prompt template of LLMs to critique and re-correct the incorrect code.}
\label{fig:method}
\end{figure}

In practice, we found that current LLMs that have undergone supervised fine-tuning (SFT) can already analyze and critique effectively. 
Therefore, we utilize in-context learning to mitigate bugs directly, rather than creating additional data for them to learn how to critique.
We design a prompt template that enables LLMs to critique and reflect on their generated code according to the programming problem and the incorrect code, identifying the locations of bugs within the code. 
Based on the root causes and locations of these bugs, we then prompt the LLMs to correct the generated code.
Additionally, we incorporate our constructed bug taxonomy and feedback from the compiler (\ie error messages from failed executions) within the prompt template, providing both holistic and individual information. 
This helps LLMs better understand the problem and the original error code, to further improve the accuracy and performance of the critique process. 
The entire process is fully automated.

The simplified prompt for our method is shown in Figure~\ref{fig:method}.
\texttt{\$DOC\_STRING} denotes the original programming problem and \texttt{\$INCORRECT\_CODE} denotes the incorrect code generated by LLMs.
\texttt{\$BUG\_CATEGORY} contains our constructed bug category and the natural interpretation of bug types.
\texttt{\$COMPILER\_FEEDBACK} is the outcome of the compiler execution.
If the result belongs to type A (\ie Syntax Bug) and type B (\ie Runtime Bug), we provide the compiler bug report; if the compiler run successfully but the code does not pass the unit tests (\ie Functional Bug), we provide the message ``The functionality of code is incorrect''; and if the execution of the code exceeds the time limit, we offer the message ``The execution of the code has exceeded the time limit''.
\texttt{\$CAUSE\_AND\_CODE} denotes the root causes and locations of bugs, and the re-corrected code generated by LLM.
The final version of the prompts is available in our artifact repository \cite{ar2024}.
The LLM used for critiquing code and fixing bugs is the same model utilized for generating code based on human requirements.
Consequently, our method eliminates the need for additional post-training costs.

We extract the re-corrected code from \texttt{\$CAUSE\_AND\_CODE} through scripts and then verify the correctness of the code by interacting with the compiler.
If the re-corrected code still does not pass unit tests, we update the prompt template with the latest incorrect code and the compiler outcome to conduct the next re-correct iteration.
In summary, our proposed self-critique-based method is cost-effective, interactive, and iterative.

\subsection{Evaluation}

\subsubsection{Effectiveness}

To evaluate the effectiveness of our method, we conducted extensive experiments across the prior nine LLMs.
For each LLM, we randomly selected 120 incorrect codes generated by the respective models, then utilized our methods of self-critique and self-revision on the code to fix bugs.
We conducted n iterations (\ie we set n = 2 in this experiment) of our method to fix bugs in each incorrect code.
The performance on GPT-4 is shown in Figure~\ref{fig:method-performance}.
The results suggest that our proposed method can effectively identify and fix bugs in incorrect code, thereby enhancing the success rate of code generation using LLMs.
Specifically, in the first iteration, GPT-4 fixes 24.1\% (29 in 120)) of incorrect self-generated code through the proposed method.
A further 6.6\% (6 in 91) of the codes are corrected in the second iteration. 
Therefore, our method achieves a repair success rate of 29.2\% (\ie (29+6)/120 = 29.2\%) after two iterations.

The syntax bugs are completely fixed in the first iteration.
For runtime bugs, 23\% of error codes are corrected in the first iteration and 16\% in the second iteration.
For functional bugs, 23\% are fixed in the first iteration and 3\% in the second iteration.
The additional experimental results for the other eight LLMs are available in our artifact repository \cite{ar2024}.

\subsubsection{Comparison Results}

To further demonstrate the effectiveness of our proposed bug repair framework, we also compare it with two popular code repair methods, \ie Self-debug \cite{chenteaching} and Intervenor \cite{wang2024intervenor}.
Self-debug employs prompt engineering to guide the LLM in repairing bugs based on compiler feedback. 
Intervenor first prompts the teacher model to generate bug repair suggestions (chain-of-thought) based on compiler feedback, and then the student model applies these suggestions to modify the code. 
All three approaches were evaluated under identical experimental settings, using the same test data and GPT-4, as shown in Table~\ref{tab:comparison-other-baseline}.

\begin{wraptable}{r}{0.5\textwidth}
  \centering
  \small
  \setlength\tabcolsep{6pt}
  \renewcommand{\arraystretch}{1.3}
  \caption{
  Comparison of methods for repairing code bugs.
  BT. Denotes Bug Taxonomy.
  }
    \begin{tabular}{l|c|c}
    \toprule
    \multicolumn{1}{c|}{\multirow{2}[4]{*}{\textbf{Method}}} & \multicolumn{2}{c}{\textbf{Repair Success Rate}} \\
\cmidrule{2-3}          & \textbf{Iteration 1} & \textbf{Iteration 2} \\
    \midrule
    Self-debug  \cite{chenteaching} & 23.3\% & 27.5\% \\
    Intervenor \cite{wang2024intervenor} & 22.5\% & 28.3\% \\
    \midrule
    \textbf{Ours} & 24.1\% & \textbf{29.2\%} \\
        \textbf{\ \ w/o BT.} & 21.7\% & 27.5\% \\
    \textbf{Ours + Intervenor} & \textbf{25.8\%} & \textbf{29.2\%} \\
    \bottomrule
\end{tabular}
  \label{tab:comparison-other-baseline}
\end{wraptable}

Experimental results indicate that our framework outperforms both self-debug and Intervenor. 
The self-debug approach achieves a lower repair success rate because it relies solely on compiler feedback for modifications. 
Although Intervenor introduces a teacher model to generate repair suggestions and a student model to apply code changes, it essentially remains dependent on compiler feedback for both explanation and modification. 
In comparison, our framework provides additional information beyond compiler results, including the potential bug type and a detailed bug explanation, which helps the LLM more accurately identify and fix bugs.

To further validate the effectiveness of our findings, we incorporated bug classification information into Intervenor’s teacher model to assess whether this additional information could improve the accuracy of repair suggestions (i.e., introducing our approach’s insights into the Intervenor framework).
The results, presented as “Ours + Intervenor,” demonstrate that our findings provide supplementary repair information, enabling Intervenor to fix more bugs.
In summary, both the findings and our framework are effective in enhancing the LLM’s understanding of bugs and facilitating bug repair.

\subsubsection{Ablation Study}

To further validate the effectiveness of bug taxonomy in our bug repair framework, we conducted ablation experiments by comparing the bug repair success rate with and without providing the bug taxonomy in the prompt.
To ensure a fair comparison, we adopted the same experimental settings. The results are presented in Table~\ref{tab:comparison-other-baseline}.
From experimental results, we draw several conclusions. 
First, the absence of bug taxonomy in the prompt led to a significant decrease in the code repair success rate. Without the bug taxonomy, the repair methods degenerate to relying solely on compiler feedback for bug localization. 
In particular, for type 3 functional bugs (constitute a significantly larger proportion of bugs, as shown in Tables~\ref{table:opensource-bugs-dv}, \ref{table:closedsource-bugs-dr} and \ref{table:rwpb}), compiler feedback alone provides limited information, typically indicating only whether the code is correct or incorrect. 
Therefore, supplementing the prompt with a checklist of possible Python bug types helps LLMs systematically consider potential errors.

Notably, when the bug taxonomy is omitted, all three methods, including self-debug \cite{chenteaching} and Intervenor \cite{wang2024intervenor}, rely only on compiler feedback for bug repair. 
However, differences in prompt design still result in performance variations among these methods. 
Finally, across two iterations, the bug repair pipelines that included the bug taxonomy consistently fixed more bugs than those without it.

\begin{figure}[t]
\centerline{\includegraphics[width=0.76\textwidth]{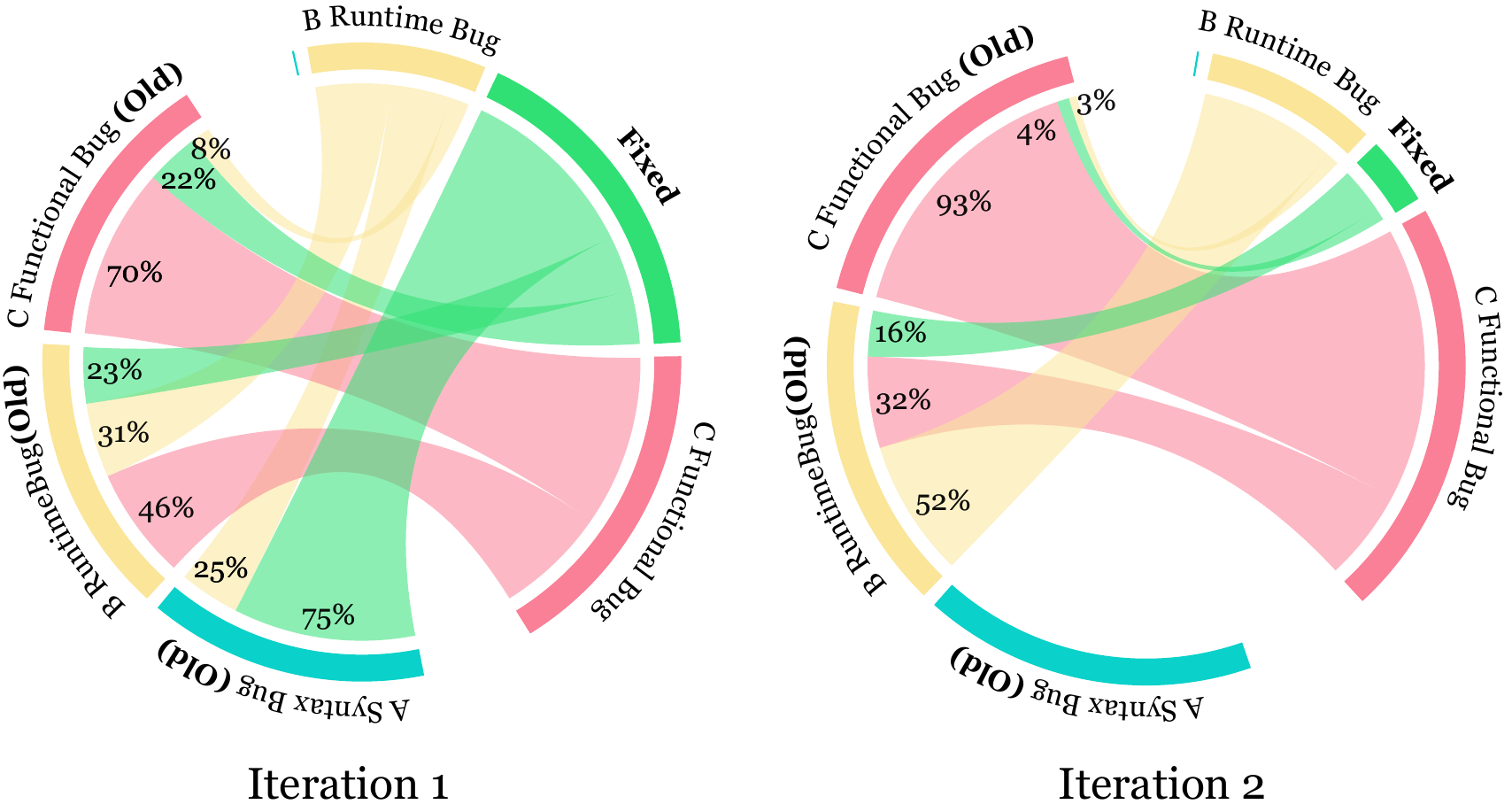}}
\caption{Results of fixing bugs by using our self-critical-based approach.}
\label{fig:method-performance}
\end{figure}

\subsubsection{Bug Evolution}

Figure~\ref{fig:method-performance} also presents the bug evolution for GPT-4.
For runtime bugs, 46\% and 32\% are transformed to functional bugs at the end of the first and second iterations, respectively.
Conversely, the functional bugs are harder to transform into the two other types of bugs.
22\% and 4\% functional bugs are fixed, and only 8\% and 3\% bugs are transformed into runtime bugs after two iterations, respectively.
Experimental results show that our approach can effectively fix all types of code to improve the performance of LLMs in code generation, while facilitating the transition of syntax and runtime bugs towards functional bugs.

In summary, in this section, we propose a bug repair framework that enables LLMs to fix their generated bugs using a self-critique mechanism.
Unlike two iterative bug repair methods we compared, our framework introduces the bug taxonomy and bug type explanations distilled from our analysis and findings.
Notably, this framework primarily serves as a straightforward application scenario for the main findings and contributions of our study.
Its purpose is to demonstrate how one aspect of our research, \ie our bug taxonomy and bug type explanations, can be applied to practical bug repair, thereby highlighting the value of our work.

Building on our study, further enhancements can be explored. 
For example:
(1) The bug taxonomy we constructed can serve as a valuable foundation for reward function design in reinforcement learning, with the potential to improve the effectiveness of reinforcement learning from compiler feedback (RLCF) approaches.
(2) Moreover, we found that incorporating information about code comments and function parameters significantly improves LLMs’ understanding of functions in real-world projects. 
Providing such information during training or at inference time may further enhance the code generation capabilities of LLMs.
(3) In Section~\ref{sec:real-world-e} and Section~\ref{sec:dis}, we detail how we automated the iterative synthesis of unit tests and function docstrings during dataset construction, and how dataset quality was ensured through manual double-checking by annotators. 
This process provides important references for future researchers aiming to automate the construction of other types of code datasets.
We hope our findings provide the community with deeper insights into LLM code generation and contribute to the development of more robust models and more efficient code repair techniques.

\section{Discussion}
\label{sec:dis}
In this section, we discuss some findings during the experiments, future work, and some significant threats to the validity of our results.

\subsection{Benchmark Analysis}

HumanEval+ and MBPP+ are widely used benchmarks in code generation with high-quality problem descriptions, diverse unit tests, and formatted input/output. 
APPS+, sourced from various algorithm online-judge websites, presents high-difficulty problems. 
However, algorithmic problems result in lengthy and often redundant descriptions (\eg extensive background to introduce the problem), complicating comprehensive understanding by LLMs. 
Furthermore, the algorithmic problems originally included images as the problem explanation, but these images are missing in the APPS+, further increasing the difficulty for models to understand the tasks.
To ensure experiment quality, we manually excluded problems with missing images.

\subsection{Annotation Process}
\label{sec:dis-anno}

During the process of analyzing and labeling bug types in incorrect code, experts are allowed to utilize GPT-4 to aid in annotation tasks. 
Specifically, for functional bugs that lack detailed compiler error information, we employ GPT-4 to generate three potential root causes with explanations, facilitating the analysis and classification of bug types.
Although the final annotations are manually conducted and subject to double-checking, we are interested in exploring the differences in effectiveness and efficiency among three annotation methods: fully automated by GPT-4, fully manual, and AI-assisted, in identifying bug types.

We randomly select 100 incorrect code samples and engage five experts who have not previously labeled these codes to separately and manually annotate the bug types without AI assistance. 
Another group of five experts uses GPT-4 for assistance to separately annotate bugs in the same manner as before.
For fully automated annotation, we provide GPT-4 with the compiler results (if have) and the bug taxonomy, and task it with analyzing the incorrect code and classifying the bug.
The temperature and top-k are 0.1 and one, respectively.
We take the double-checked results as ground truths and report the average accuracy of these three annotation methods, as shown in Table~\ref{tab:dis-anno}.
The results show that, on secondary categories, the accuracy of fully automated annotation stands at 37\%.
This indicates that GPT-4 can provide coarse-grained bug categories of bugs to aid expert annotation.
For tertiary categories, the accuracy of AI-assisted, fully manual, and fully automated is 91.6\%, 84.4\%, and 5\%, respectively.
It indicates that AI-assisted annotation enhances accuracy compared to fully manual methods, which is a conclusion similar to the findings of Saunders et al. \cite{saunders2022self} on the topic-based summarization task.
Moreover, fully automated annotation is not yet able to replace manual annotation to achieve fully automated annotation in our task.
On the other hand, AI-assisted can enhance the efficiency of the annotation process.
The experts provide feedback that annotating with the help of bug causes given by the GPT-4 can expedite their labeling tasks.

\begin{table}[htbp]
  \centering
  \small
  \setlength\tabcolsep{6pt}
  \renewcommand{\arraystretch}{1.0}
  \caption{The average accuracy of three annotation methods (\ie AI-assisted, fully manual, and fully automated) in identifying bug types.}
\begin{tabular}{c|ccc}
\toprule
\textbf{Bug Category Level} & \textbf{AI-Assisted} & \textbf{Fully Manual}  & \textbf{Fully Automated} \\
\midrule
Secondary Category & 98.2\% & 99.0\% & 37\% \\
\midrule
Tertiary Category & 91.6\% & 84.4\% & 5\% \\
\bottomrule
\end{tabular}
  \label{tab:dis-anno}
\end{table}

\subsection{Automating \emph{RWPB} Construction Process}
\label{sec:dis-auto}

In Section~\ref{sec:real-world-bencmark-cons}, we propose a pipeline to manually construct a real-world benchmark, while trying our best to prevent data leakage.
However, an ideal situation would be to automate the construction of high-quality datasets. 
The automated process can easily synthesize large amounts of data that can be used not only for evaluation, but also to expand the size of our training data, increasing the ease of use of the method. 
In this section, we discuss the strategy for automating our manual construction process.

The initial \emph{RWPB} build process can be summarized into five parts:
(1) start with real-world code repositories, filtering for those that are recently released and popular;
(2) use automated scripts to filter and clean the functions extracted from these repositories;
(3) remove duplicate functions using both open-source and closed-source large-scale code corpora;
(4) manually synthesize docstrings, including programming problem descriptions, arguments, and returns;
(5) generate unit test examples and ensure they cover the entire code comprehensively.
In fact, the first, second, and third steps, as well as the calculation of unit test coverage in the fifth step, can all be automated through scripts. 
Therefore, we focus on two key questions: \emph{in generating docstrings and unit tests, the difference in performance between automated and manual methods.}

\begin{table}[htbp]
  \centering
  \small
  \setlength\tabcolsep{6pt}
  \renewcommand{\arraystretch}{1.3}
  \caption{The average results in synthesizing compliant docstrings.}
\begin{tabular}{c|ccc|c}
\toprule
\textbf{Docstring} & \textbf{Naturalness} & \textbf{Accuracy}  & \textbf{Clarity} & \textbf{All Met} \\
\midrule
Programming Problem Description &    97.6\% &     92.1\% &  90.9\%   &     87.3\% \\
\midrule
Arguments and Returns & - & 76.4\% & - & - \\
\bottomrule
\end{tabular}
  \label{tab:dis-auto-1}
\end{table}

To assess the feasibility of automation, we design two experiments.
First, for the canonical code of all samples in \emph{RWPB}, we ask GPT-4 to generate the corresponding docstrings (\ie the programming problem descriptions, arguments, and returns) according to the three criteria we used for annotation (\ie naturalness, accuracy, and clarity).
The prompt for generating docstrings is available in our artifact repository \cite{ar2024}. 
We then assign three experts to separately evaluate whether the docstrings generated by GPT-4 met these criteria.
Table~\ref{tab:dis-auto-1} presents the average ratio of compliant docstrings. 
From the results, we found that GPT-4 has a 97.6\%, 92.1\%, and 90.9\% success rate in synthesizing problem descriptions that meet the ``naturalness'', ``accuracy'', and ``clarity'' criteria, respectively.
The proportion of generating problem descriptions that satisfy all criteria is 87.3\%.
These results show that GPT-4 exhibits high accuracy in single-function comprehension and can provide an accurate problem description based on the canonical code. 
However, GPT-4 has room for improvement in accurately generating the functions' arguments and returns. 
The success rate for accurately synthesizing arguments and returns decreases to 76.4\%.
For instance, GPT-4 struggles to accurately generate nested types and more complex types such as Tensors.
Therefore, enhancing the LLM's accuracy in understanding and generating function arguments and returns may become a future key research issue for automated code data synthesis.

\begin{table}[htbp]
  \centering
  \small
  \setlength\tabcolsep{6pt}
  \renewcommand{\arraystretch}{1.3}
  \caption{The average success rate in synthesizing compliant unit test inputs according to the canonical code and the docstring.}
\begin{tabular}{c|cc}
\toprule
\textbf{Docstring Source}  & \textbf{At Least One Execution Successful} & \textbf{All Executions Successful} \\
\midrule
GPT-4 Generated Docstring &    69.2\% &     66.9\%  \\
\midrule
Human Written Docstring & 88.6\% & 85.7\% \\
\bottomrule
\end{tabular}
  \label{tab:dis-auto-2}
\end{table}

Second, we evaluate GPT-4's performance in generating inputs for unit tests given the canonical code and the docstring. 
We task GPT-4 to generate diverse unit test inputs, including regular and boundary cases, to cover the entire code.
The docstrings are categorized into those previously generated by GPT-4 and those manually written in \emph{RWPB}. 
The unit test output can be obtained by compiling and executing the canonical code.
The experimental results are shown in Table~\ref{tab:dis-auto-2}.
From the results, we found that when provided with human-written docstrings, the total execution success rate of GPT-4's synthesized unit tests is 85.7\%. 
However, when the docstrings contain inaccuracies, particularly incorrect arguments and return types, the total success rate of the synthesis decreases to 66.9\%. 
This indicates that more accurate information must be provided to LLMs when using them to automatically generate unit tests.
Additionally, we found that in the initial generation, some test samples did not achieve full coverage of the canonical code, and some raised compilation errors. 
To address this issue, we provide the unexecuted code segments or compilation error messages in the prompt to GPT-4 to iteratively generate additional unit tests. 
The results show that GPT-4 can improve the unit test coverage iteratively.

In summary, while GPT-4 ensures acceptable accuracy in automatically synthesizing programming problem descriptions, there is still room for improvement in the precision of generating the function's arguments and returns.

\subsection{Future Work}

Our study used a fixed prompt template to guide LLMs in completing programming problems. 
This approach allows us to control experimental settings and directly investigate the types of bugs that LLMs might introduce during code generation. 
However, various prompts and demonstrations (\ie in-context learning) can improve LLM's performance on the code generation task. 
In the future, we plan to explore the differences in bugs introduced by LLMs when using various prompts compared to our used prompt, examining aspects such as variations in bug categories and their proportional distributions.
Additionally, we ensure the uniqueness of the generated code for the same problem by setting the temperature to 0.1 and using a top-k sampling with k set to 1. 
In the future, we plan to investigate the effect of different decoding strategies on bug distributions.

In this study, all benchmarks we used were monolingual, meaning the canonical code was written in Python. 
Although recent efforts have focused on enhancing LLMs' code generation capabilities in Python, and the results also on these benchmarks are indicative of their overall code generation proficiency, in future work, we plan to extend our methodology to multilingual scenarios, further analyzing the types and causes of bugs in code generated by LLMs across different programming languages.

We discussed a way to automate our dataset-building process, as shown in Section~\ref{sec:dis-auto}. 
We plan to develop a more effective code data synthesis pipeline that will assist not only in the evaluation process but also in the training process in the future.
Finally, we introduced a self-critique-based method to enable LLMs to fix the bugs they generate. 
In the future, we plan to further leverage our bug categorization and experimental findings to enhance the performance of code generation and self-based bug fixing.

\subsection{Threats to Validity}

\textbf{Dataset Validity.}
The first concern involves potential issues with the popular benchmarks we used.
APPS+ contains a wide range of programming problems with varying difficulty, which can thoroughly test the code generation capabilities of LLMs. 
However, there are certain issues with this benchmark. 
First, the problem descriptions often include a lot of background information irrelevant to the core algorithm, which can interfere with the LLM's understanding of the problem. 
Furthermore, some descriptions use images for explanations, but these images are merely represented by ``[image]'', which prevents the LLM from fully comprehending the problem. 
To avoid these issues, we manually checked APPS+ and carefully removed cases with inaccurate or incomplete descriptions.

Another issue is that all benchmarks we used were monolingual, \ie the programming language of the canonical code was in Python.
Our study focuses exclusively on Python rather than conducting experiments across multiple programming languages for several reasons: (1) current popular LLMs demonstrate their strongest code generation abilities in Python \cite{bai2023qwen,roziere2023codellama}.
Therefore, analyzing model performance on Python provides a reliable measure of code generation capabilities and allows for a thorough understanding of potential bugs \cite{chen2021evaluating}.
(2) Python programming benchmarks are currently the most comprehensive, including HumanEval+ \cite{chen2021evaluating}, MBPP+ \cite{austin2021program} and APPS+ \cite{apps} (these benchmarks also being those which we used in our study).
These benchmarks facilitate in-depth analysis of LLMs’ code generation abilities and enable detailed examination of bug types and characteristics. 
This is essential for understanding LLMs’ code generation performance and for broadening the scope of future research.
(3) Code generation across multiple programming languages is a distinct research topic.
Our study aims to provide a comprehensive understanding of LLMs’ code generation abilities and bug characteristics in Python, both in experimental and real-world settings.
Including multiple programming languages would introduce considerable complexity and dilute the focus and clarity of our work. 
Moreover, the diversity of programming language syntax, semantics, and ecosystem-specific bugs would require a fundamentally different experimental design, which is beyond the scope of this study. 
Therefore, we deliberately focus on Python to ensure a rigorous and systematic investigation.
In the future, we plan to investigate and analyze limitations faced by LLMs in generating other programming languages.

The final issue is that the benchmarks may cause data leakage \cite{jain2024livecodebench, golchin2023time, weller2023according, riddell2024quantifying, roberts2023cutoff, shi2023detecting, zhou2023don} as current LLMs are trained on a large fraction of GitHub, StackOverflow and code libraries \cite{chen2021evaluating}.
In our process of benchmark construction, we collected samples from the latest code repositories and utilized a de-duplication algorithm (\ie MinHash \cite{broder1997resemblance}) on code training corpora including open-source bases (\ie RedPajama Data v2, Stack v1 \cite{kocetkovstack} and v2 \cite{lozhkov2024starcoder2}, RefinedWeb \cite{penedo2023refinedweb}, and Pile \cite{gao2020pile}) and our internal code training corpora, to mitigate data leakage.

\textbf{External Validity.}
Threats to external validity pertain to the generalizability of our results across different scenarios. 
To address these concerns, we evaluated nine state-of-the-art LLMs, comprising four open-source textual LLMs, two open-source code LLMs, and three closed-source LLMs.
We employed three established benchmarks and developed a real-world benchmark to assess LLM performance in practical code generation tasks. 
Our experimental setup follows protocols from prior research, supporting the generalizability of our findings to other contexts.

\textbf{Internal Validity.}
When labeling bug types in code generation, we use manual categorization with the assistance of GPT-4. 
This approach can improve the accuracy of the process but may cause ``lazy annotation'' (details in Section~\ref{sec:dis-anno}). 
Additionally, different experts may have varying understandings of bug types, which potentially might have resulted in labeling errors. 
To ensure consistency, we employ double-checking among different experts. 
This method helps us achieve a consensus on bug types, avoiding incorrect labeling.
Additionally, because LLMs may generate varying responses for the same problem, we standardize the generation process by setting the temperature to 0.1 and using top-1, following established practices \cite{gpt4tr, Guo2024DeepSeekCoderWT}.

\section{Related Work}

\textbf{Large Language Models for Code Generation.}
Recently, the advancement of LLMs has significantly propelled the field of code generation.
Due to training on large code corpora, several text LLMs have demonstrated remarkable ability in understanding natural language and synthesizing code, as seen in models such as Mistral \cite{jiang2023mistral}, Qwen1.5 \cite{bai2023qwen},  DeepSeek-LLM \cite{deepseek-llm}, the ChatGPT family \cite{gpt35, gpt4tr}, the Claude3 family \cite{anthropic2024claude}, and the Llama family \cite{llama2, llama3}.
On the other hand, code LLMs, primarily trained on extensive code data including CodeGPT \cite{lu2021codexglue}, SantaCoder \cite{allal2023santacoder}, CodeGeex \cite{zheng2023codegeex},  CodeX \cite{chen2021evaluating}, WizardCoder \cite{luo2023wizardcoder}, CodeLlama \cite{roziere2023codellama}, DeepSeekCoder \cite{Guo2024DeepSeekCoderWT}, Phi-3-Instruct \cite{abdin2024phi}, and StarCoder-2-Instruct \cite{starcoder}, have also exhibited remarkable capabilities in code generation.
Researchers have further improved the performance of LLMs on code generation by prompt engineering \cite{le2023codechain, shin2023prompt, denny2023conversing} or by introducing supervised fine-tuning \cite{jain2023llm, bairi2023codeplan, weyssow2023exploring} and reinforcement learning \cite{dou2024stepcoder, yu2023mathcal, le2022coderl, liu2023rltf, shojaee2023execution, zheng2023secrets}.

\textbf{Bug Study in Software Engineering.}
Investigating bugs in code generation helps developers understand the limitations of models and improve their performance.
Previous research has analyzed bugs introduced by deep learning models and frameworks \cite{islam2019comprehensive, wang2022empirical, zhang2018empirical, garcia2020comprehensive}, as well as by LLMs in various coding tasks \cite{pan2024lost, xia2022less, yang2024exploring, liu2024exploring, mousavi2024investigation, tambon2024bugs}. 
In contrast to these studies, our study focuses on bugs introduced by LLMs.

We are the second study to comprehensively investigate the types of bugs that can be introduced by code generated by LLMs, following the work of Tambon et al \cite{tambon2024bugs}.
However, Tambon et al. \cite{tambon2024bugs} focus on bugs in LLM-generated non-standalone code, which differs from our study.
Secondly, they conducted all experiments on only one benchmark and three LLMs, which poses a threat to the validity and generalizability of their results and conclusions.
In contrast, we conducted our experiments on three widely used benchmarks, one real-world benchmark, and nine popular LLMs (\ie four open-source text LLMs, two open-source code LLMs, and three closed-source text LLMs).
Thirdly, Tambon et al. categorized bugs into 10 broad categories, which we found to be quite coarse.
Our categorization is more detailed, consisting of three major categories, ten secondary categories, and additional tertiary categories within each secondary category.
Finally, they also overlooked the differences in bug distribution between real-world projects and existing benchmarks, as well as the potential data leakage that may occur in existing benchmarks.
In contrast, we elaborately constructed a real-world benchmark while actively working to avoid data leakage.
We comprehensively investigated the effectiveness of LLMs on code generation for real-world projects and analyzed LLM-generated bugs in real-world scenarios.
Moreover, based on our findings, we also introduced a self-critique-based method to mitigate the bugs in LLM-generated code.
In summary, our study significantly differs in terms of the types, scenarios, and sources of programming problems we focus on, as well as in our bug taxonomy and analysis perspectives. 
Our study encompasses a wider range of benchmarks and LLMs.
Additionally, we manually constructed a benchmark of real-world scenarios, analyzed the differences in bugs within this context, and proposed a novel self-repair method based on our findings.

\textbf{Code Generation Benchmarks.}
Several works have proposed benchmarks to evaluate the capability of LLMs in code generation.
Some benchmarks are constructed from code exercises and introductory programming problems \cite{chen2021evaluating, austin2021program, liu2024your}, or completion questions \cite{apps, dai2024mhpp}.
However, these benchmarks are far removed from real-world program development.
For instance, Liu et al. \cite{liu2024your} proposed to automatically generate additional unit test samples for existing benchmarks (\ie HumanEval and MBPP) to enhance their reliability.
In contrast, we aim to construct a new high-quality real-world benchmark while making every effort to prevent data leakage to evaluate current LLMs. 
We also discuss the feasibility of automating our manual dataset construction process in Section~\ref{sec:dis-auto}.

Additionally, some benchmarks are collected from the popular public libraries and code repositories \cite{zancert, wang2022execution}, which may cause data leakage due to significant overlap with the datasets used for training LLMs \cite{starcoder, chen2021evaluating}.
Furthermore, several researchers aim to introduce non-standalone \cite{iyer2018mapping, yu2024codereval}, repository-level \cite{liu2023repobench, li2024deveval}, and GitHub issue \cite{jimenez2023swe} benchmarks.
In contrast to these benchmarks, our proposed \textit{RWPB} aims to evaluate the code generation capability of LLMs in real-world projects and standalone scenarios.
Moreover, during the filtering phase of constructing \textit{RWPB}, we de-duplicated data with large open-source and our private code corpora to avoid data leakage.
We also evaluated the distributions of bugs introduced by LLMs based on our constructed bug taxonomy, which further helps us understand how they perform in real-world code generation scenarios.

\section{Conclusion}

In this paper, we conduct a comprehensive empirical investigation into the effectiveness and limitations of code generation using LLMs.
We first conduct extensive experiments from multiple perspectives on nine widely used LLMs across three popular benchmarks. 
Our results reveal that these LLMs struggle to generate accurate code for those more complex problems.
Subsequently, we manually annotate bug types, construct a taxonomy of these bugs, analyze their distributions, and summarize 14 findings leading LLMs to generate erroneous code. 
To evaluate the effectiveness of LLMs in real-world projects, we design a rigorous benchmark construction process to minimize data leakage and construct a real-world project benchmark, \textit{RWPB}.
We analyzed the distributions of bugs on \textit{RWPB} to help researchers understand how they perform in real-world code generation scenarios.
Finally, to mitigate bugs during code generation by LLMs, we propose a novel method that introduces self-critique, enabling LLMs to iteratively critique their generated code and fix bugs.

\Acknowledgements{
The authors would like to express their sincere gratitude to Weikang ZHOU, Caishuang HUANG, Yan LIU, Enyu ZHOU, Yuhao ZHOU, Rongxiang WENG, Jingang WANG, and Xunliang CAI for their invaluable contributions to this study.
The authors extend their appreciation to the annotators for their dedicated support throughout this research.
The authors also wish to thank the anonymous reviewers for their insightful and constructive comments.
This work was partially funded by National Natural Science Foundation of China (No.62476061, 62376061, 62206057, 62372193, U2436207), Shanghai Rising-Star Program (23QA1400200), Natural Science Foundation of Shanghai (23ZR1403500), Program of Shanghai Academic Research Leader under grant 22XD1401100.
}

\end{document}